\documentclass[a4paper,11pt,nofootinbib]{revtex4}
\usepackage[frenchb,english]{babel}
\usepackage[applemac]{inputenc}
\usepackage{enumerate}
\usepackage{amsmath}
\usepackage{amsthm}
\usepackage{amssymb}
\usepackage{graphicx}
\usepackage{floatflt}
\usepackage{fancybox}
\usepackage{pict2e}
\usepackage{color}
\usepackage{hyperref}

\begin{document}

\renewcommand{\Re}{\text{Re}}
\renewcommand{\Im}{\text{Im}}

\newcommand{\om} \omega   
\newcommand{\Om} \Omega
\newcommand{\eps} \epsilon
\newcommand{\la} \lambda

\newcommand{\tal} {\tilde\alpha}
\newcommand{\tga} {\tilde \gamma}

\newcommand{\tT}{\tilde T}
\newcommand{\tR}{\tilde R}

\newcommand{\be} {\begin{equation}}
\newcommand{\ee} {\end{equation}}

 \newcommand{\ba} {\begin{eqnarray}}
 \newcommand{\ea} {\end{eqnarray}}

\newcommand{\bal} {\begin{align}}
\newcommand{\eal} {\end{align}}

\newcommand{\sba}{\begin{subeqnarray}}
\newcommand{\sea}{\end{subeqnarray}}

\newcommand{\Eqref}[1]{Eq.~(\ref{#1})}
\newcommand{\ket}[1] {\mbox{$ \vert #1 \rangle $}}
\newcommand{\kett}[1] {\mbox{$ \vert #1 \rangle_2 $}}
\newcommand{\bra}[1] {\mbox{$ \langle #1 \vert $}}
\newcommand{\abs}[1] {\mbox{$ \vert #1 \vert $}}
\newcommand{\vect}[1]{\overrightarrow{#1}}
\newcommand{\inv}[1]{\frac{1}{#1}}
\newcommand{\dpartialt}{\stackrel{\leftrightarrow}{\partial_t}}
\newcommand{\dpartial}{\stackrel{\leftrightarrow}{\partial}}
\newcommand{\dpartialeta}{\stackrel{\leftrightarrow}{\partial_{\eta}}}
\newcommand{\av}[1]{\langle\!\langle \,  #1 \, \rangle\!\rangle}
\newcommand{\intk}{\int\!\!\frac{d^3k}{(2\pi)^{3/2}}\,}
\newcommand{\tprod}{\widetilde{\prod\limits_{\bold k}}}
\newcommand{\tprodl}{\widetilde{\prod\limits_{\lambda}}}
\newcommand{\tintk}{\int\!\!\frac{d^3\tilde k}{(2\pi)^{\frac{3}{2}}}\,}
\newcommand{\tint}{\int\!\!d^{3}\tilde k\,}

\def\lrD{\mathrel{{\cal D}\kern-1.em\raise1.75ex\hbox{$\leftrightarrow$}}}
\def\lr #1{\mathrel{#1\kern-1.25em\raise1.75ex\hbox{$\leftrightarrow$}}}

\selectlanguage{english}

\baselineskip 17pt

\title{Black hole radiation with short distance dispersion, \\
an analytical S-matrix approach}

\author{Antonin Coutant}
\email{antonin.coutant@th.u-psud.fr}
\affiliation{Laboratoire de Physique Th\'eorique, CNRS UMR 8627, B\^at. 210, Universit\'e Paris-Sud 11, 91405 Orsay Cedex, France}
\author{Renaud Parentani}
\email{renaud.parentani@th.u-psud.fr}
\affiliation{Laboratoire de Physique Th\'eorique, CNRS UMR 8627, B\^at. 210, Universit\'e Paris-Sud 11, 91405 Orsay Cedex, France}
\author{Stefano Finazzi}
\email{finazzi@sissa.it}
\affiliation{SISSA, via Bonomea 265, Trieste 34151, Italy and INFN sezione di Trieste, Via Valerio 2, Trieste 34127, Italy}

\date{\today}

\begin{abstract}
Local and non-local 
properties of Hawking radiation in the presence of short distance dispersion are 
computed using connection formulae. The robustness of the spectrum and that of the two-point function 
are explained by showing that the leading deviations from the 
relativistic expressions decrease with the inverse of the spatial extension of the near horizon region. 
This region corresponds to a portion of de Sitter space with a preferred frame.
We show that the phases of the Bogoliubov coefficients are relevant 
for the two-point function in  black and white holes, and also for the black hole laser effect. 
We also present an unexpected relation between the spectra obtained 
with sub and with superluminal dispersion and we apply our formalism to massive fields. 
Our predictions are validated by numerical analysis. 

\end{abstract}

\maketitle

\newpage

\section*{Introduction}
Hawking radiation (HR), the spontaneous and steady emission of thermal radiation by black holes~\cite{Hawk75}, 
plays a crucial role in black hole thermodynamics~\cite{BCH72,Bek73,Bek74} and in the attempts to 
construct theories of quantum gravity~\cite{CallanMald,TJthermo}. Given that these are still 
incomplete, it is of importance to determine under which conditions HR occurs, 
and how its properties depend or not on hypothesis concerning the ultra-violet behavior of the theory.
In fact, because modes with arbitrary short wavelengths are involved in the standard 
derivation~\cite{Unruh81,Primer,tHooft85},
it was questioned~\cite{TJ91,TJ93} whether this process would still be present if Lorentz invariance were broken on very short scales. This question was motivated, on one hand, by the {\it possibility} that  
quantum gravitational effects could be effectively described by a non trivial dispersive relation, see {\it e.g.}~\cite{beyond07}, and, on the other hand, by the {\it fact} that phonons, or other collective degrees of freedom, obey dispersive equations at short distance, {\it e.g.} when approaching the inter atomic scale.  Hence dispersion must be taken into account when computing the phonon spectrum that a black hole analogue~\cite{Unruh81} would emit. To this end, Unruh~\cite{Unruh95} wrote a dispersive wave 
equation in an acoustic black hole metric. He then numerically found that the thermal properties of the flux are robust, {\it i.e.} not significantly affected when $\Lambda \gg \kappa$, where $\Lambda$ is the 
ultra-violet dispersive frequency, and $\kappa$  the surface gravity of the black hole. 
In a subsequent numerical analysis~\cite{CJ96}, it was observed that ``{\it the radiation is astonishingly close to a perfect thermal spectrum}''. 
This was confirmed to a higher accuracy in~\cite{Macher1}, and partially explained by analytical treatments~\cite{BMPS,Corley,Tanaka99,SU,Rivista05}.

In spite of these works, the origin of this astonishing robustness is not completely understood. This is due to our ignorance of the parameters 
governing the first deviations with respect to the standard thermal spectrum. 
In the present work, we show that when $\Lambda \gg \kappa$ the most relevant parameter is the extension 
of the near horizon region in which second order gradients can be 
neglected.~\footnote{\label{f1} When these higher derivatives vanish, 
the background corresponds to a de Sitter space endowed with an homogeneous 
preferred frame that respects some of the de Sitter isometries. 
This was noticed in~\cite{Constructing07}, and further developed 
with Jean Macher during his PhD~\cite{Macherthesis}.
This was exploited in~\cite{From2010}, and we hope to report on it soon.}
We reach these results by studying the validity limits 
of the connection formula~\cite{BMPS,Corley,Tanaka99,SU} encoding the 
scattering across the horizon. While the core of the calculation is based on the mode 
properties near the horizon  -- which are universal as they rely on a first order expansion around 
the horizon-- the validity limits are essentially governed by the extension of the 
region where this expansion is valid. 
As a result, on the one hand, the leading order expressions are universal and 
agree with the standard relativistic ones, and on the other hand, the first deviations depend on this extension. 
Their evaluation is carried out using a superluminal dispersion relation. Interestingly, these results also apply to subluminal dispersion, as we show in establishing a correspondence 
between these two cases.
We apply our analysis to both black and white holes, and show that HR in the latter backgrounds
gives rise to undulations~\cite{Mayoral2011,SilkePRL2010} that contribute 
as classical waves to the observables.
We then compare our predictions with the numerical works~\cite{FP2,FP3} which were done in parallel 
with the present analysis. We apply our treatment to the 
black hole laser effect in App.~\ref{BHL} and to massive fields in App.~\ref{massApp}. 

\section{Black hole metrics and dispersive theories}
\subsection{The choices of settings}
\label{settings}

To study the propagation of a dispersive field on a black hole geometry, basically two lines of thought can be followed. First one can study a particular fluid and derive the equations for linearized perturbations from the known microscopic theory,
 {\it e.g.} of a Bose condensate~\cite{MacherBEC}. 
Secondly, more abstractly, one can identify the ingredients that must be adopted in order to obtain a well-defined mode equation. We adopt here this second attitude~\cite{Unruh95,BMPS,TJ96} 
as it is more general, and as it reveals what are exactly the choices that should be made.

First one needs to choose a dispersion relation in a flat (homogeneous) situation
\be
\Omega^2 = F^2(p), \label{disp}
\ee
where $\Omega$ is the frequency in the reference frame, $p$ is the norm of the wave vector, and $F^2  = p^2 \pm p^4/\Lambda^2 + o(p^4)$. We thus work in units so that the group velocity is unity for $\Omega \to 0$. The critical frequency $\Lambda$ sets the scale at which dispersive effects become significant. In what follows, to simplify the equations 
we shall often work with
\be
F^2(p) = \left(p + \frac{p^3}{2 \Lambda^2} \right)^2 = p^2 + \frac{p^4}{\Lambda^2} + \frac{p^6}{4\Lambda^4} \label{dispr}.
\ee
At the end of the paper we shall consider more general relations. 

Second one needs to choose the two background fields, namely the metric $g_{\mu \nu}$, and the `preferred' frame, {\it i.e.} the {\it local} frame in which eq. (\ref{disp}) is implemented. As explained in~\cite{TJ96} the latter 
can be described by a unit time-like vector field, $u^\mu$, in terms of which the 
preferred frequency is $\Omega = u^\mu p_\mu$. In what follows, for reasons of simplicity, we work in 1+1 dimensions. 
Then the 'preferred' spatial wave vector $p$ is $p = s^\mu p_\mu$, where $s^\mu u_\mu = 0$, 
and $s^\mu s_\mu = -1$.  
For further simplicity, we also work with stationary background fields, and we impose that the 
flow $dx^\mu/d\tau = u^\mu$ is geodesic. Then, 
the Painlev\'e-Gullstrand coordinates~\cite{AnalogueLivingReview},
are the 'preferred' coordinates as they obey $dt = u_\mu dx^\mu$
and $\partial_x = s^\mu \partial_\mu$. Using them, 
$g_{\mu \nu}$ and $u_\mu$ are both encoded by a single function $v(x)$: 
\be
ds^2 = dt^2 - (dx - v(x) dt)^2 \label{metric},
\ee
and 
\be
\Omega = \omega - v(x) p, \label{Om}
\ee
where $\omega$ is the conserved frequency associated with the Killing field $\partial_t$. 
Returning to the first attitude above discussed, both Eq.~(\ref{metric}) and Eq.~(\ref{Om}) 
are obtained in the lab coordinates when considering the dispersive sound propagation 
characterized by Eq.~(\ref{disp})
in a fluid whose flow velocity is $v(x)$.

When $v^2(x)$ crosses 1, there is a Killing horizon. More precisely, for $v < 0$, it is a black (white) 
hole horizon when $\kappa = \partial_x v$ evaluated at the horizon is positive (negative). This gradient determines the surface gravity of the horizon: neglecting dispersion, {\it i.e.} when $F^2 = p^2$, the near horizon behavior of null right moving (upstream) geodesics is 
\be
x = x_0 \ e^{\kappa t}.
\label{xkt}
\ee
Correspondingly, since $xp = cst$ near the horizon, 
the redshift experienced by right moving wave packets is~\cite{BMPS,Rivista05}
 \be
 p = p_0 \, e^{-\kappa t}. \label{pexp}
 \ee

The third choice concerns the quantization procedure. There exist indeed many inequivalent wave equations associated 
with the dispersion relation
\be
 (\omega - v p)^2 = F^2(p) \label{disp2}.
 \ee
This remain true even when using an action to guarantee that the wave equation be governed by self-adjoint differential operators. In this paper we shall work with~\cite{Unruh95} 
\be
\left[ \left( \omega + i \partial_x v \right) \left( \omega + i v \partial_x \right)  -  F^2(i\partial_x) \right] \varphi_\omega = 0 \label{modequ},
\ee
which is nothing but
\be
\left[ \left( \partial_t + \partial_x v \right) \left( \partial_t + v \partial_x \right)  +  F^2(i\partial_x) \right] \phi = 0 \label{wavequ},
\ee
applied on a stationary mode $\phi = e^{- i \omega t} \varphi_\omega$. The associated conserved scalar product is
\be
(\phi_1|\phi_2) = i\int_{\mathbb R} \left[ \phi_1^*(\partial_t + v\partial_x)\phi_2 - \phi_2 (\partial_t + v\partial_x) \phi_1^*\right] dx.\label{scalt}
\ee
Furthermore, we work with profiles $v(x)$ defined on the entire real axis, and 
asymptotically constant: $|v(\pm \infty)| < \infty$. Hence
the domain of integration is the entire real axis. 
The stationary (positive norm and real frequency) modes $\phi_\omega$ are then normalized by
\be
(\phi_{\omega'}|\phi_\omega) = \delta(\omega' - \omega).
\label{scalom}
\ee
In conformity with second quantization~\cite{Waldbook}, negative norm modes are named $(\phi_{-\omega})^*$, 
so that $\phi_{-\omega}$ are positive norm modes of negative frequency $-\omega$. 

Eq.~(\ref{wavequ}) reduces to the scalar massless d'Alembert equation when $F^2 = p^2 = - \partial_x^2$. 
It differs in several respects from the (two dimensional) Bogoliubov-de Gennes equation~\cite{MacherBEC}.
It first differs in the hydrodynamical regime, {\it i.e.} in the limit $\Lambda \to \infty$, in that the latter equation is not conformally invariant. As a result, left and right moving modes remain coupled in that case. Moreover, it also differs from Eq.~(\ref{modequ}) when taking into account the quartic dispersive effects. The differences arise from different orderings of $\partial_x$'s and $v(x)$. 
To give another example, Eq.~(\ref{modequ}) also differs from alternative models~\cite{BMPS,SU08} in which left and right moving modes remain exactly decoupled even when $F(p)$ is non linear. 

Nevertheless, these wave equations share the same characteristics since these 
are determined by \Eqref{disp2}. 
What is less obvious is that these models also share, at leading order, the same 
deviations of the spectrum which are due to dispersion. 
This follows from the fact that these deviations
are based on asymptotic expansions that are 
governed by Hamilton-Jacobi actions
associated with \Eqref{disp2}.

\subsection{{\it in} and {\it out} mode basis, connection formulae, and Hawking radiation}
\label{bogo}

\begin{figure}[!h]
\begin{center} 
\includegraphics[scale=1]{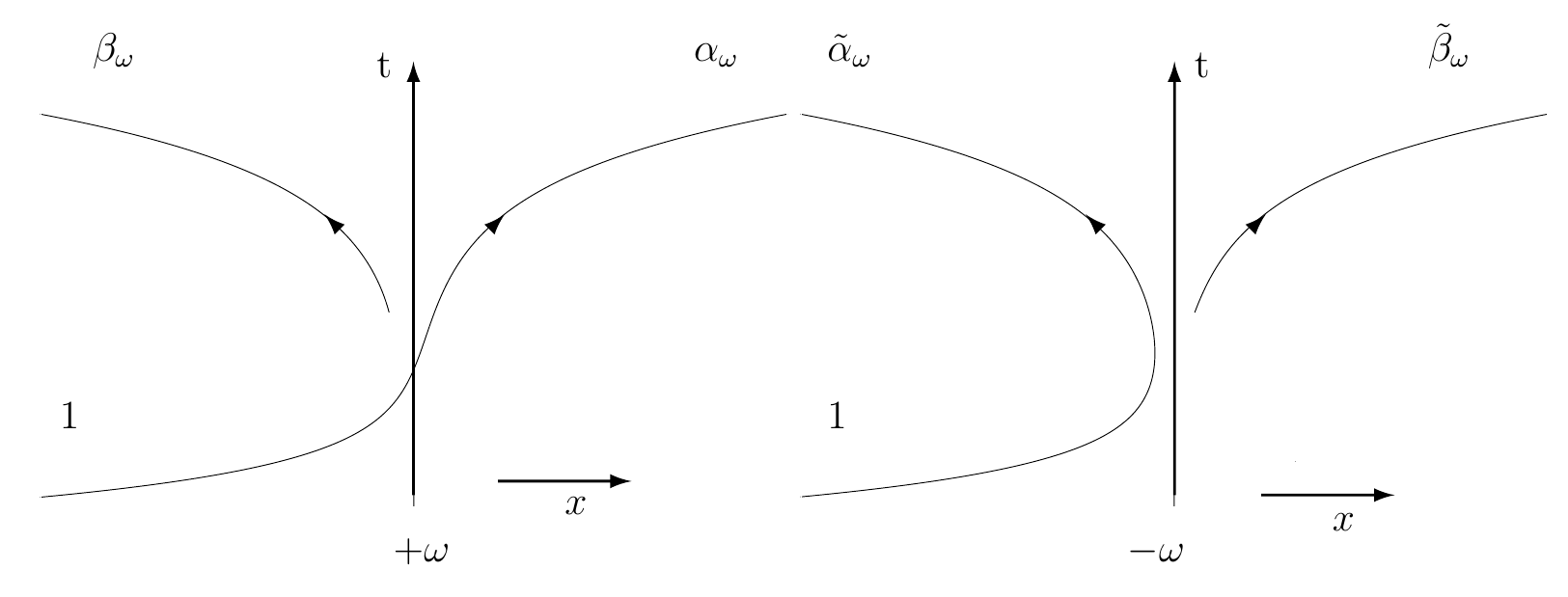}
\end{center}
\caption{Space-time representation of the near horizon trajectories followed by wave packets made 
with right moving (with respect to the fluid) 
{\it in} modes in the presence of superluminal dispersion, {\it e.g.} that given in \Eqref{dispr},
and in the coordinate system $(t,x)$ of \Eqref{metric}.
We have also indicated their asymptotic amplitudes as given in Eq.~(\ref{BogHR}). 
On the left panel we have represented the trajectories  
associated with $\phi_\omega^{\rm in}$, and on the right those of the partner mode  $(\phi^{\rm in}_{-\omega})^*$
with negative norm. The turning point of the latter is given in \Eqref{xtp}.} 
\label{BogoBH} 
\end{figure}

In this paper, the properties of HR 
will be approximately determined by making use of connection formulae that relate 
asymptotic solutions of Eq.~(\ref{modequ}).
Before describing this 
procedure in precise terms, let us briefly explain it. 
For the stationary profiles we consider, {\it i.e.} with $v$ defined on the entire real axis and asymptotically constant, because of dispersion, the Bogoliubov transformation encoding the Hawking effect has the standard form of a scattering matrix. It should be stressed that this is not the case 
for relativistic fields. In that case indeed, because of \Eqref{xkt}, wave packets propagated backwards in time hug onto 
the horizon for arbitrary long time, and thus never transform into waves incoming from an asymptotic region. 
Instead, when there is dispersion, \Eqref{xkt} is followed only for a finite
time, and wave packets (propagated backwards in time) leave the near horizon region and reach, for superluminal dispersion,
$x = -\infty$, see Fig.~\ref{BogoBH}. 

Moreover when wave packets reach the asymptotic
regions where $v$ is constant, they can be decomposed in terms of stationary plane waves $e^{- i \om t }e^{i p_\om x}$.
Hence, the definition of the {\it in} and {\it out} modes is the standard one, see {\it e.g.} the scattering in a constant electric field~\cite{Primer}. The {\it in} modes $\phi_\omega^{\rm in}$ are solutions of \Eqref{modequ} such that the group velocity $v_{\rm gr} = (\partial_\om p_\om)^{-1}$ of their 
asymptotic branches is oriented towards the horizon for one of them only. Hence when 
forming a wave packet of such modes, it 
initially describes a single packet traveling towards the horizon,
whereas at later time it describes several packets moving away from the horizon. Similarly,
the {\it out} modes $\phi_\omega^{\rm out}$ contain only one asymptotic branch with a group velocity oriented 
away from the horizon.

At this point it should be noticed that the dimensionality of these two sets depends 
on the asymptotic values of $v$. 
When $v(x)$ contains one horizon, {\it i.e.} crosses $-1$ once, the dimensionality is $3$ 
below a threshold frequency $\om_{\rm max}$~\cite{Macher1}: for $ \om_{\rm max} > \om > 0$, 
there is one positive norm left moving mode (with respect to the fluid, but 
not the lab) $\phi^{\rm left}_\omega$, and a pair of right moving ones of opposite norm that we shall call 
$\phi_\omega$ and $(\phi_{-\omega})^*$ according to the sign of their norm. 
Hence the scattering matrix is $3 \times 3$. 
However, when $v$ is smooth enough, 
$\phi^{\rm left}_\omega$ essentially decouples. This has been numerically shown in~\cite{Macher1}, and is mathematically justified in 
App.\ref{appWKB}.
Hence to a very good approximation, one recovers a 
$2 \times 2$ matrix characterizing right moving modes only. From now on we shall work within this approximation. 

Introducing the {\it in} and {\it out} sets of modes, they are related by
\be
\begin{pmatrix} \phi_\omega^{\rm in} \\ (\phi_{-\omega}^{\rm in})^* \end{pmatrix} =\begin{pmatrix} \alpha_\om & \beta_\om \\ \tilde \beta_\om & \tilde \alpha_\om \end{pmatrix} \cdot \begin{pmatrix} \phi_\omega^{\rm out} \\ (\phi_{-\omega}^{\rm out})^* \end{pmatrix} \label{BogHR}.
\ee
Because starred modes have a negative norm, the  
matrix is an element of $U(1,1)$.
That is, the coefficients obey:
\ba
|\alpha_\om|^2 - |\beta_\om|^2 &= 1, \nonumber \\
\alpha_\om^*\tilde \beta_\om - \beta_\om^*\tilde \alpha_\om &= 0, \label{scatrel}
\ea
and $|\beta_\om|^2 =  |\tilde \beta_\om|^2 $. Hence, when working in the {\it in} vacuum, the mean number of emitted pairs of quanta of opposite frequency is 
\be
\bar n_\om = \vert \beta_\om \vert^2.
\label{e13}
\ee
In the relativistic limit, 
{\it i.e.} $\Lambda\to\infty$, one gets the standard result
\be
\bar n_\om^{\rm relativistic} = \frac{1}{e^{2 \pi \om/\kappa} - 1},
\label{relats}
\ee
{\it i.e.}, a Planck distribution at the Hawking temperature $T_{\rm H} = \kappa/2\pi$, in units where $\hbar = 1 $ and $
k_B = 1$.

Our aim is to compute the coefficients of \Eqref{BogHR} using a connection formula. To this end, we  
first identify the various asymptotic solutions, and then we evaluate 
the globally defined solutions which we match to the asymptotic ones.
These techniques have been already used in~\cite{BMPS,Corley,Tanaka99,SU}. 
The novelty of our treatment concerns a careful control of the various 
approximations involved in this procedure. This will enable us to 
control the leading deviations from \Eqref{relats} which are due to dispersion, given $v(x)$.

\subsection{The relevant properties of the profile $v$}
\label{profileSec}

To be able to compute these leading deviations  it is necessary to further discuss the properties of the profile $v(x)$. When using relativistic fields, the 
temperature of HR is completely fixed by $\kappa = \partial_x v$, the gradient of $v$ evaluated at the horizon, 
even though the asymptotic flux generally depends on other properties of $v(x)$ which fix the `grey body factors'~\cite{Hawk75,Page76}. However these describe an elastic scattering between $\phi^{\rm left}_\om$ and $\phi_\om$, 
and therefore do not affect the temperature, as can be verified by considering the equilibrium state described by the Hartle-Hawking vacuum~\cite{BirrelDavies,Primer}. 
When dealing with dispersive fields, 
$\kappa$ is no longer the only relevant quantity. Indeed, as we shall show in the sequel, {\it several} properties of $v(x)$ now become relevant. Moreover they govern 
{\it different} types of deviations with respect to the standard flux. For {\it smooth} profiles, there are basically two important properties,
near horizon ones, and asymptotic ones. 

If there is a regular horizon at $x=0$, $v$ can be expanded as
\be
v(x) =-1 + \kappa x + O(x^2). \label{vlin}
\ee
This near horizon behavior is only valid in a certain interval, not necessarily symmetric about $0$. 
Hence we define $D_{\rm lin}^{L}$ and $D_{\rm lin}^{R}$ such that for
\be
- D_{\rm lin}^{L} \lesssim \kappa x \lesssim D_{\rm lin}^{R}, \label{Dlin}
\ee
$v$ is linear, to a good approximation (see region 1 in Fig.~\ref{regions}). 
As we shall later establish, $D_{\rm lin}^{L}$ and $D_{\rm lin}^{R}$ 
control the leading deviations of the spectrum.
It is worth noticing that in the limit $D_{\rm lin}^{L}, D_{\rm lin}^{R} \to \infty $, 
one effectively works in de Sitter space {\it with} a preferred frame since \Eqref{Om} and \Eqref{disp2} 
still apply. In that limiting case, as we shall explain,
the relativistic result of \Eqref{relats} is found with a higher accuracy.

The other relevant parameter is related to the
asymptotic values of $v$, that we assume to be finite. For superluminal dispersion, 
what matters is 
\be
v(x = - \infty) = -1 - D_{\rm as} < -1 \label{Das}.
\ee
For subluminal instead, it is $v (x = +\infty)$ that matters.
As discussed in~\cite{Macher1}, 
$D_{\rm as}$ determines the critical frequency $\omega_{\rm max}$ 
(computed below in \Eqref{ommax})
above which the flux vanishes~\footnote{In that work, the profile was $v= - 1 + D \tanh(\kappa x/D)$. Hence, when looking at the deviations of the spectrum upon changing $D$,  the deviations associated with $D_{\rm lin}$ and $D_{\rm as}$ were confused since both scaled in the same way. In fact, one of the main novelty of the present analysis, and the companion numerical works~\cite{FP2,FP3}, 
is to remove this confusion by analyzing the deviations due to $D_{\rm lin}^R$ and $D_{\rm lin}^L$ only.}.

For finite values of $x$, $v$ can have a complicated behavior. As we said, we only suppose that the interpolation between the asymptotic regions and around the horizon is {\it smooth} enough, so we can neglect non-adiabatic effects. Indeed, a sharpness in 
$v(x)$ induces non-adiabatic effects~\cite{CJ96,Macher1,FP2} 
not related to the Hawking effect that both destroy the thermality of the spectrum and induce 
higher couplings between left and right moving modes, see Figs. 12 and 16 in \cite{Macher1}. 
These effects are due to the breakdown of the WKB approximation studied in App.\ref{appWKB}.

\subsection{Hamilton-Jacobi actions and turning point}
\label{HJSec}
\subsubsection{Hamilton-Jacobi actions and turning point}
\label{trajcla}

\Eqref{disp2} can be interpreted as the Hamilton-Jacobi equation of a point particle. Indeed it suffices to consider the solution for $p$ as a function of $x$ at fixed $\omega$, which we call $p_\omega(x)$, as $p_\omega(x) = \partial_x S_\omega$ where the action is decomposed as $S = - \omega t + S_\omega(x)$. One thus works the standard expression
\be
S_{\omega}(x) =\int^x  p_{\omega}(x') dx'.
\ee
In these classical terms, left and right moving solutions with respect to the fluid, 
{\it i.e.} the roots of $\omega - v p = \pm  F$, 
decouple and can be studied separately. 
Restricting attention to the right moving ones which enter in \Eqref{BogHR}, 
one deals with 
\be
\omega - v(x) p_\omega(x) =  F(p_\omega(x)).
\label{rDR}
\ee
In the sequel, it will be useful to work in the $p$-representation with 
\be
W_\omega(p) = - px + S_\omega(x) =  - \int^p  X_{\omega}(p') dp' \label{Wom}.
\ee
In this representation, at fixed $\omega$, the position $x$ is viewed as a function of $p$. It is given by $X_\omega = \partial_p W_\omega$ 
and obeys
\be
\omega - v(X_\omega(p))\ p =  F(p) .
\label{Xomp}
\ee
The usefulness of the $p$-representation can be appreciated when considering 
the trajectories in near the horizon region, where $v \sim - 1 + \kappa x$. 
In this region, irrespectively of $F(p)$, one finds
\be
{dp \over dt} = -\left( \frac{\partial X_\omega}{\partial \omega} \right)^{-1} = -\kappa p,
\ee
which gives \Eqref{pexp}, as in relativistic settings. Then the trajectory 
is algebraically given by
\be \label{Xp}
x_F(t) = X_\omega(p(t)) = \frac{\omega}{p \kappa} - \frac{F(p) - p}{\kappa p}.
\ee
Unlike what is found for $p(t)$, using \Eqref{dispr}, 
 $x_F$ completely differs from \Eqref{xkt} for $|p|  \geq \Lambda$. 
To adopt a language appropriate to the study of the modes, we shall work with $\om > 0$ only.
Then negative frequency roots $p_{-\om} > 0$ of \Eqref{rDR}
are replaced by the negative roots $p_\om< 0$ associated with $\om > 0$, as explained in~\cite{TJ96}. 
Hence, \Eqref{Xp} defines two trajectories, one with $p > 0$, and one with $p < 0$. At early times, {\it i.e.} for large $|p| \gg \omega $ 
and for superluminal dispersion $|F/p| > 1$, both are 
coming from the supersonic region $x< 0$.  
Then, for $p > 0$ the trajectory 
crosses the horizon and reaches $x = +\infty$, 
whereas for $p  < 0$, it is reflected back to $x = -\infty$, see Fig.\ref{BogoBH}. 
What is important is that 
both trajectories stay in the near horizon region a finite time $\Delta t$. 
The integrated red-shift effect $p_{final}/p_{initial} = e^{-\kappa \Delta t}$ is thus finite, 
unlike what is found for relativistic propagation
where \Eqref{pexp} applies to arbitrary early times. 

For $p < 0$, the location of the turning point $x_{tp}$ is obtained by solving 
$d x/dt = (\partial_\omega p_{\omega})^{-1} = 0$. Using Eq.~(\ref{dispr}), 
\Eqref{rDR} gives
\be
\omega = (1+v) p_\omega + \frac{p_\omega^3}{2\Lambda^2}. 
\label{xHJ}
\ee
Hence, the momentum and the velocity at the turning point are (exactly) 
\be
- p_{tp}= \left(\omega \Lambda^2 \right)^{\frac13},
\label{ptp}
\ee
and 
\be
v(x_{tp})+1 = - \frac{3}{2} \left(\frac{\omega}{\Lambda} \right)^{\frac23} \label{vtp}.
\ee
If $\omega$ is sufficiently small, {\it i.e.} $\omega < \Lambda \left( D^L_{\rm lin}\right)^{\frac32}$,
the turning point is located in the near horizon region, and it is given by
\be
\kappa x_{tp} = - \frac{3}{2} \left(\frac{\omega}{\Lambda} \right)^{\frac23} \label{xtp}.
\ee

\begin{figure}[!h]
\begin{center} 
\includegraphics[scale=1]{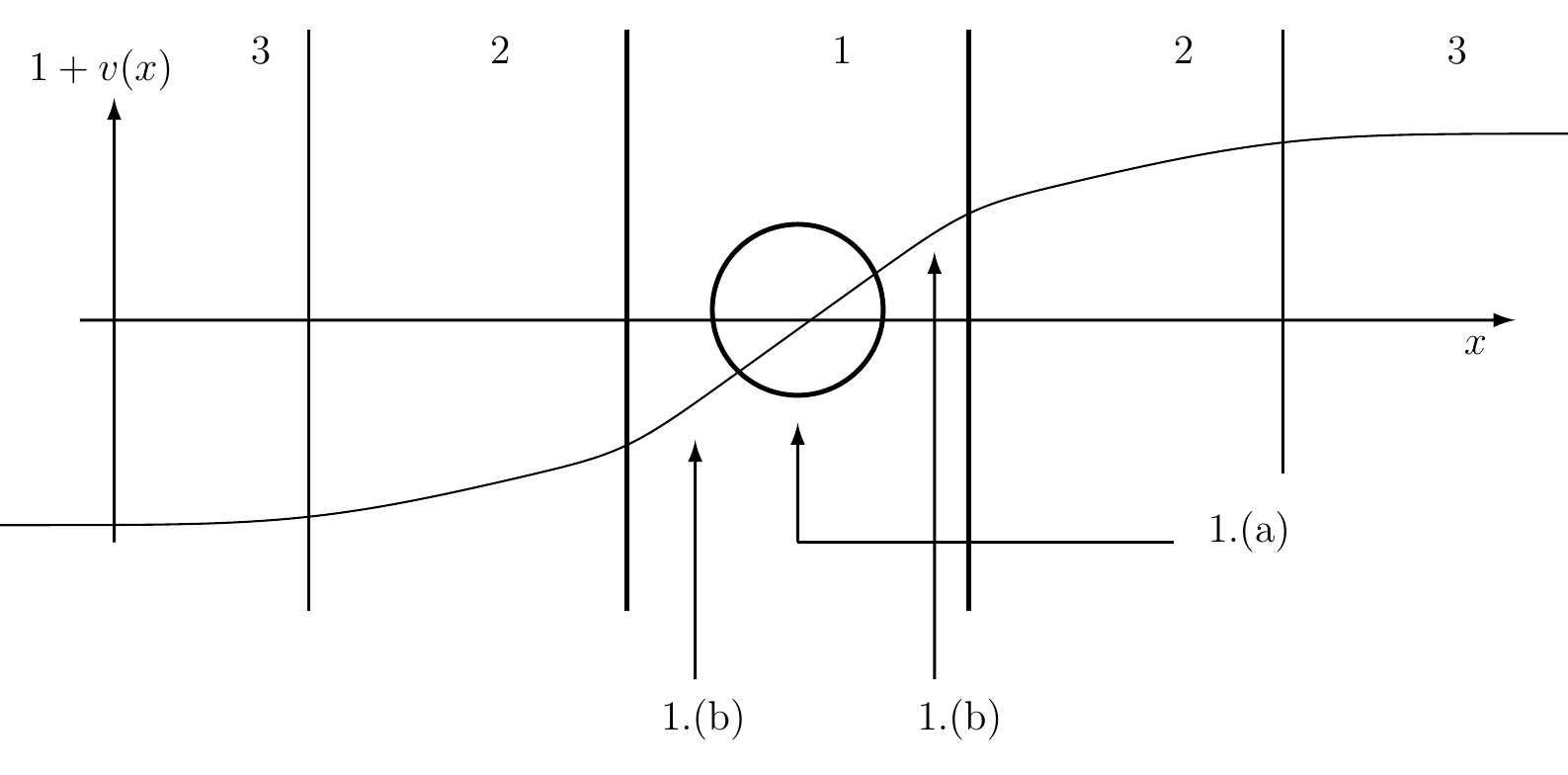}
\end{center}
\caption{Shape of the typical velocity profile $v$, 
together with the extension of the relevant regions. For a given value of $\omega$, 
the near horizon region $1$ where $v \sim -1 + \kappa x$ splits into two: 
a region close to the turning point of \Eqref{xtp} (1.(a)) 
where the WKB approximation fails, and two intermediate regions (1.(b)) where 
this approximation becomes reliable, and whose sizes
are fixed by $D_{\rm lin}^{L}$ and $D_{\rm lin}^{R}$. 
In the asymptotically flat regions 3, the solutions are superpositions of plane waves. 
The intermediate regions 2 play no significant role when $v$ is sufficiently smooth, 
because the propagation is accurately WKB, {\it i.e.} no mode mixing.
As we shall see, mode mixing essentially occurs at the scale of the turning point, {\it i.e.} in 
region 1.(a).}
\label{regions} 
\end{figure}

In classical terms, the main consequence of dispersion 
is the introduction of this turning point. It 
introduces a non trivial multiplicity of the real roots $p_\omega(x)$, 
solutions of Eq.~(\ref{xHJ}). This multiplicity will play a crucial role when solving the mode equation~(\ref{modequ}). From \Eqref{vtp} and \Eqref{Das}, 
we see that there is no turning point for $\om$ above 
\be
\omega_{\rm max} = \Lambda \left(\frac23 D_{\rm as} \right)^{\frac32} \label{ommax}.
\ee
This is threshold frequency $\omega_{\rm max}$ mentioned in Sec.\ref{bogo}. 
It corresponds to the limiting case where the turning point $x_{tp}$ is sent to $- \infty$. 
For $\om > \omega_{\rm max} $ only 
the positive root of \Eqref{xHJ} and the positive norm mode $\phi_\om$ remain. 
Thus the transformation of \Eqref{BogHR} no longer exists.

While these results have been obtained with a superluminal dispersion relation, however, they hold when the dispersion is subluminal.
Indeed \Eqref{xHJ} is invariant under the three replacements: 
\begin{align}
1+v &\rightarrow -(1+v),\nonumber \\
\omega &\rightarrow -\omega, \label{subsuper} \\
\frac1{\Lambda^2} &\rightarrow -\frac1{\Lambda^2}. 
\nonumber 
\end{align}
The first replacement 
exchanges the subsonic region and the supersonic one (for $v< 0$). 
As a result, a black hole horizon is replaced 
by a white one and {\it vice versa}.
The second one amounts to a time reverse symmetry, $t \to - t$. 
At the classical level, it exchanges the roles of positive and negative 
roots of \Eqref{disp2}. At the mode level, it changes the sign of their norm, as discussed in Sec.~\ref{bogo}. 
The third replaces a superluminal quartic dispersion by a subluminal one. This exchange
applies to all dispersion relations when expressed as $F(p)- p \to -(F(p)-p)$. It replaces 
any dispersion that exhibits a superluminal character when $p$ approaches $\Lambda$ 
by the corresponding subluminal one.
This correspondence thus 
applies to dispersion relations that pass from super to sub, as it is the case gravity waves in water
when taking into account capillary waves~\cite{Rousseaux}.

Under \Eqref{subsuper}, the trajectories are mapped into each other, as the function $X_\om(p)$ of \Eqref{Xomp} is {\it unchanged}. 
Hence Eqs. (\ref{ptp},\ref{vtp},\ref{xtp}), are also
unaffected. Because we changed the sign of $\om$, for subluminal dispersion,
it is the trajectory of positive frequency that has a turning point. 

\subsubsection{The action $S_\om$ in the near horizon region}
\label{actionSec}
In preparation for the mode analysis we study the behavior of $S_\omega$ 
in the near horizon region where $v$ is linear in $x$. Because of this linear character, 
it was appropriate to first solve the equations of motion in the $p$-representation and then go in 
the $x$-representation. This is also true for the action itself. 
Moreover, when solving \Eqref{modequ}, $\phi_\om(x)$ will be obtained by inverse Fourier transforming the mode in $p$-space $\tilde \phi_\om(p)$. 
Thus we express the action as $S_\om = xp - W_\om(p)$. 
Imposing $\partial_p S = 0$, we get
\be
S_{\omega}(x,p_0) = x p_\omega(x) - \int_{p_0}^{p_\omega(x)} X_\omega(p) dp \label{Somint},
\ee
where $p_\om$ is a solution of \Eqref{xHJ} and $p_0$ fixes the integration constant. 
Using \Eqref{Xp} and \Eqref{dispr}, one gets
\be
S_{\omega}(x,p_0) = x p_{\omega}(x) - \frac{\omega}{\kappa} \ln \left (p_{\omega}(x)\right) + \frac{(p_{\omega}(x))^3}{6\Lambda^2 \kappa} + \theta_0,
\label{Som}
\ee
where $\theta_0$ is 
\begin{align}
\theta_0 = \frac{\omega}{\kappa} \ln \left(p_0 \right) - \frac{p_0^3}{6\Lambda^2 \kappa}.
\label{theta0}
\end{align}
To consider all solutions of \Eqref{modequ}, we shall compute the 
action for all roots of \Eqref{xHJ}, including the complex ones. 
To this end, we need to define the integral $\int^p_{p_0} dp'/p' = \ln (p/p_0)$, that arises 
from the first term of \Eqref{Xp}, for $p$ complex. We shall work with the argument of cut equal to $\pi - \epsilon$,
so that $\ln(-1) = -i\pi$. 

\section{Mode analysis}
\label{modeanalys}
\subsection{Asymptotic mode basis and $x$-WKB approximation}
\label{xmodeanalys}
Since the WKB approximation fails near a turning point, we cannot compute
the coefficients of \Eqref{BogHR} using this approximation. In fact, under this approximation one 
would get a trivial result, namely $\beta_\om = \tilde \beta_\om = 0$, 
$\vert \alpha_\om \vert = \vert \tilde \alpha_\om \vert = 1$. 
To get a non-trivial result, 
 we shall compute these coefficients by inverse Fourier 
transforming the modes in $p$-space and identifying the various terms sufficiently far away 
from the turning point, in a calculation that generalizes that of the Airy function~\cite{AbramoSteg}.

In App.\ref{appWKB} we present the calculation of the WKB modes of \Eqref{wavequ} 
which generalizes the usual treatment since \Eqref{wavequ} is not second order. We also evaluate the errors made with respect to the exact solutions. 
In particular, the relevant 
errors are {\it bounded} by the inverse of the dimensionless parameter
\be
\Delta(x) = \frac{\Lambda}{2\kappa}|2(1+v(x))|^{\frac32} \label{Delta}.
\ee
As expected, far away from the horizon $\Delta$ is large and the WKB approximation is 
accurate. More precisely 
$\Delta$ becomes of order 1 near $x=x_{tp}$ of \Eqref{xtp} evaluated for $\omega \sim \kappa$. Hence 
for these frequencies, which are the relevant ones for HR, 
$\Delta \gg 1$ is reached for $x/x_{tp} \gg 1$. 
One also sees that at fixed $x$,
$\Delta$ grows like $\Lambda/\kappa\to \infty$, hence $\Delta$ also governs the dispersion-less limit.

\subsubsection{The six roots $p_\omega$ far away from the turning point}
\label{approxHJ}

Since we work in a weak dispersive regime, {\it i.e.} $\Lambda \gg \kappa$, and since HR is related to 
frequencies $\omega \sim \kappa$, we have $\omega \ll \Lambda$ for relevant frequencies. 
Moreover, since we impose that $D_{\rm as}$ is not too small, we also 
have $\omega \ll \omega_{\rm max}$, see \Eqref{ommax}.
Hence $\omega_{\rm max}$ only concerns the high frequency properties of HR, which we no longer study. 
We focus instead on frequencies $\omega \sim \kappa$. Even for such frequencies, the expressions of $p_{\omega}(x)$, solutions of Eq.~(\ref{xHJ}), are 
complicated. However, their exact expression is not needed.
It is sufficient to estimate them far away from the turning point, in order to build the 
mode basis. 

\begin{figure}[!h]
\begin{center} 
\includegraphics[scale=0.6]{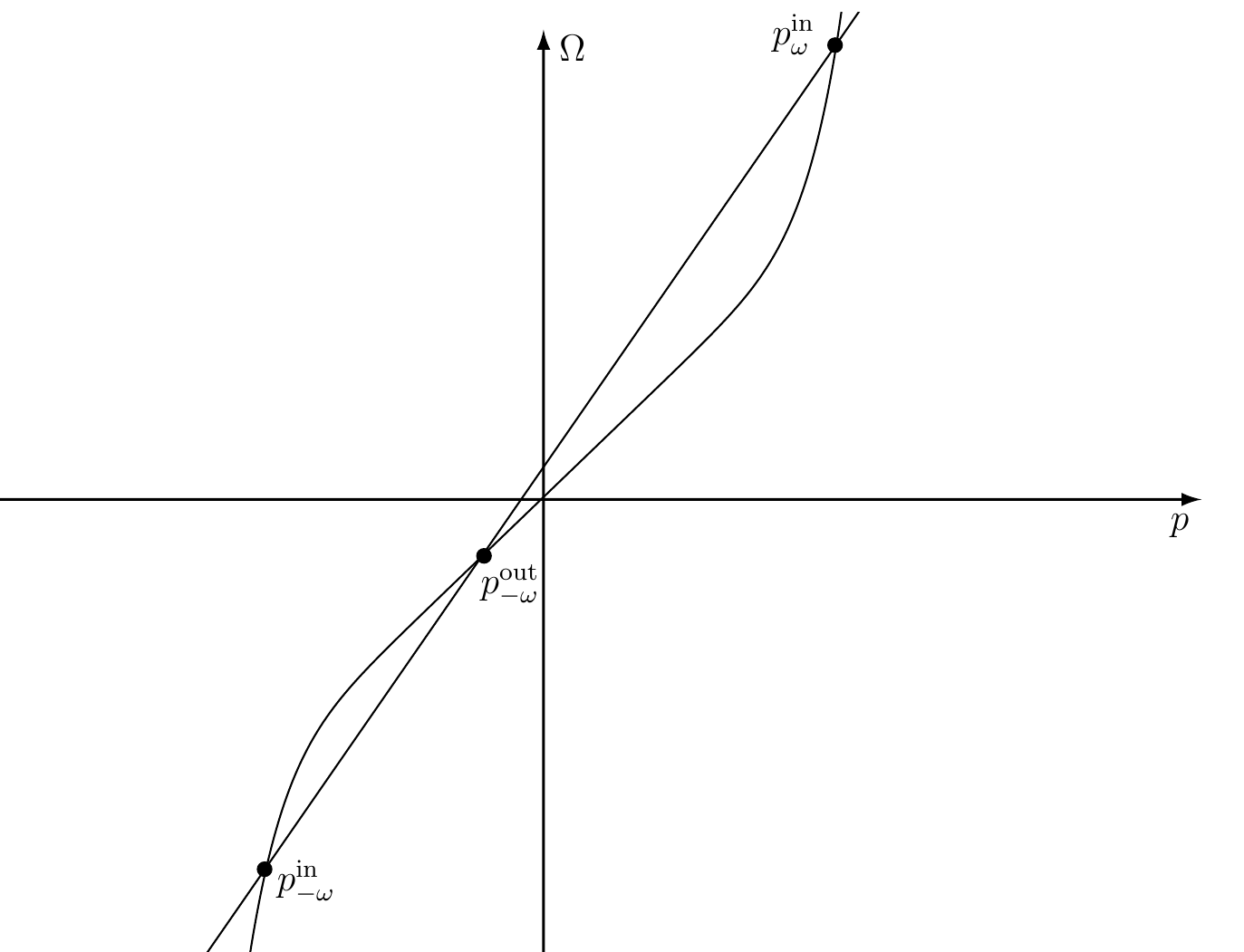}
\includegraphics[scale=0.6]{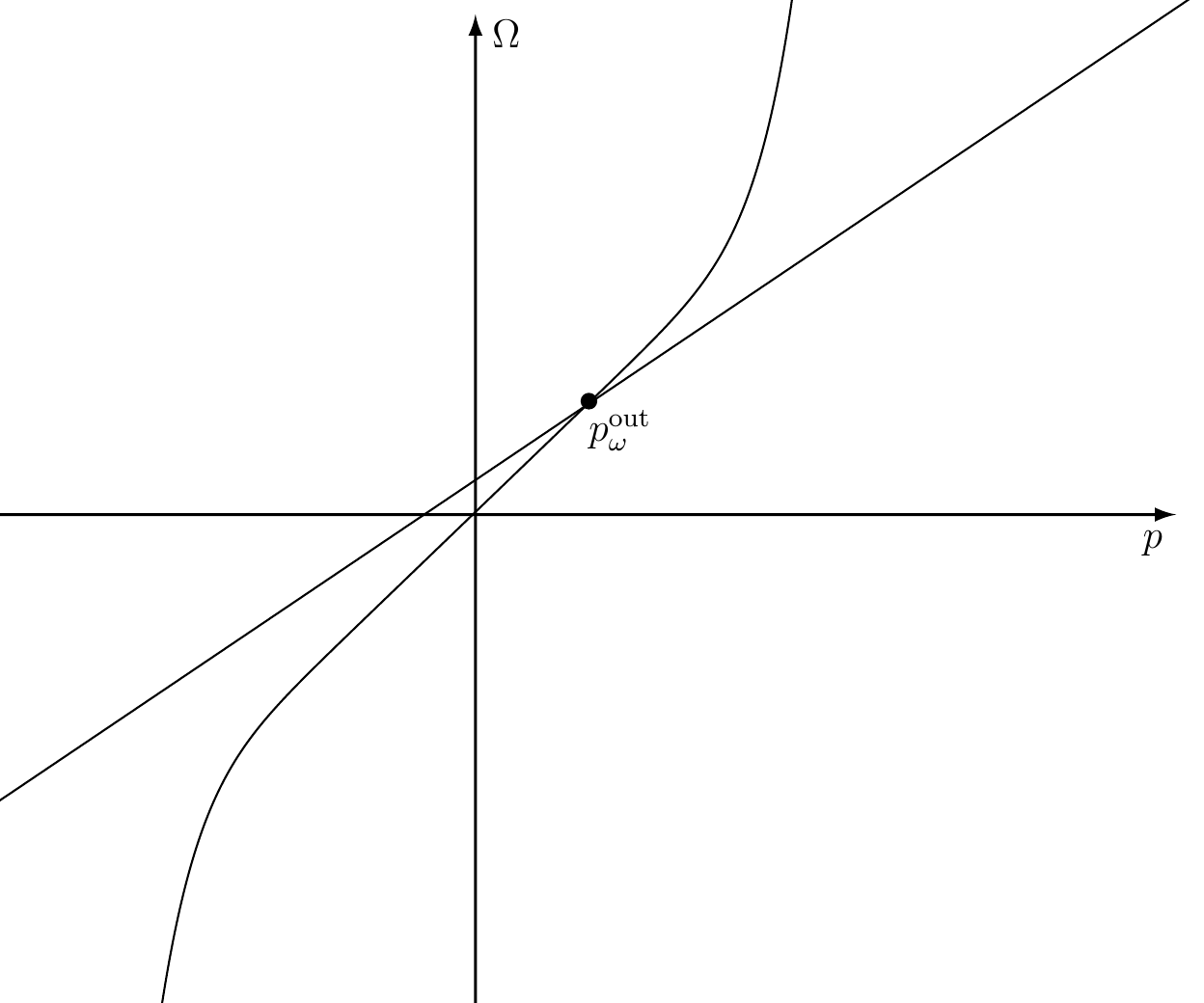}
\end{center}
\caption{Graphical resolution of \Eqref{xHJ} restricted to right moving modes.
The dispersion relation $F = p + p^3/2\Lambda^2$ and 
the straight lines $\Om = \om - v p$ are plotted for a common value of $\om$ and two 
different values of $v$. On the left side, $1+v<0$ and we have 3 real roots given by Eqs. (\ref{plusinroot}, \ref{minusinroot}, \ref{minusoutroot}). On the right side, $1+v>0$ and we have the single real root of \Eqref{plusoutroot}. The two complex roots of Eqs. (\ref{grroot}, \ref{decroot}), are not represented.} 
\label{graphroots} 
\end{figure}

The denomination of the roots we use is based on that of the corresponding mode,
which is itself based on the sign of the group velocity, as explained in Sec.\ref{bogo}. 
Moreover, we exploit the fact that for right moving modes 
the sign of the norm is that of the corresponding root $p_\om$ (see {\it e.g.}~\cite{CJ96}). 
Therefore, for $\om > 0$, positive roots shall be called $p_\omega$, whereas 
negative roots shall be called $p_{-\omega}$ in accord with the fact that 
negative norm modes are called $\left( \phi_{-\om}\right)^*$, 
see Fig.~\ref{graphroots}. 
Using this terminology, on the left of the turning point, one finds 
\begin{align}
p_\omega^{\rm in} &= \Lambda \sqrt{-2(1+v)} - \frac{\omega}{2(1+v)} (1+ O(y)), \label{plusinroot}\\
p_{-\omega}^{\rm in} &= -\Lambda \sqrt{-2(1+v)} - \frac{\omega}{2(1+v)} (1+ O(y)), \label{minusinroot}\\
p_{-\omega}^{\rm out} &= -\frac{\omega}{1+ v} (1+ O(y^2)), \label{minusoutroot}
\end{align} 
where the small parameter $y$ is related to $\Delta$ of \Eqref{Delta} by 
\be
y = \frac{ \omega/\kappa}{\Delta(x)}. 
\label{y}
\ee
Far away from the turning point of \Eqref{xtp}, $y \ll 1$ and our expressions are reliable approximations.
On the right (subsonic) region, one has only one real root 
\be
p_\omega^{\rm out} = \frac{\omega}{1+ v} (1+ O(y^2)).\label{plusoutroot}
\ee
On this side there are also two complex solutions which do not correspond to any classical trajectory. 
However, when looking at the solutions of \Eqref{modequ}, 
they govern the growing mode $\phi_\om^{\uparrow}$ and the decaying mode
$\phi_\om^{\downarrow}$ exactly as real roots govern WKB modes. These roots are 
\begin{align}
p_\omega^{\uparrow} &= - i\Lambda \sqrt{2(1+v)} - \frac{\omega}{2(1+v)} (1+ O(y)),\label{grroot}\\
p_{\omega}^{\downarrow} &= i\Lambda \sqrt{2(1+v)} - \frac{\omega}{2(1+v)} (1+ O(y)). \label{decroot}
\end{align}
On this side of the horizon, the corrections are again governed by $y$ of Eq.~(\ref{y}). 
Hence, for $\om \sim \kappa$, the corrections to the roots are on both sides 
controlled by $\Delta$ of \Eqref{Delta}. It should be also noticed that
the errors for the two {\it out} roots are subdominant with respect 
to the other ones.

\subsubsection{The six WKB modes far away from the turning point}
\label{basis}

In order to distinguish globally defined modes from their WKB approximations, 
the former shall be noted $\phi_\om$, and the latter $\varphi_\om$. 
Since we consider only $\om>0$, negative frequency modes (of negative norm) 
shall be written as $\left( \phi _{-\om}\right)^*$ or $\left( \varphi _{-\om}\right)^*$.

Sufficiently far away from the turning point, {\it i.e.} for $\Delta \gg 1$, 
the WKB modes offer reliable solutions of \Eqref{modequ}. 
They can thus be used as a basis to decompose globally defined solutions. 
We also assume that $D_{\rm lin}^{R,L}$ of \Eqref{Dlin}  
are large enough so that one can be at the same time be far away from the turning point
and still in the region where $v$ is linear. 
Using the expressions for the six roots in this region, 
\Eqref{Som}, and neglecting the common phase depending on $\theta_0$ of \Eqref{theta0},  
on the left side of the horizon one obtains 
\begin{align}
\varphi_\omega^{\rm in} &\sim
[2\Delta(x)]^{-\frac12} \frac{e^{-i\frac{\omega}{\kappa} \ln(\Lambda \sqrt{2\kappa})}}{\sqrt{4\pi \kappa(1+ \kappa |x|)}} |x|^{-i\frac{\omega}{2\kappa}} e^{- i \frac23 \Delta(x)}, \label{plusin}\\
\left( \varphi_{-\omega}^{\rm in} \right)^* &\sim
[2\Delta(x)]^{-\frac12} \frac{e^{-i\frac{\omega}{\kappa} \ln(\Lambda \sqrt{2\kappa})}}{\sqrt{4\pi \kappa(1+ \kappa |x|)}} |x|^{-i\frac{\omega}{2\kappa}} e^{i \frac23 \Delta(x)},\label{minusin}\\
\left( \varphi_{-\omega}^{\rm out}\right)^* &\sim e^{i\frac{\omega}{\kappa} -i \frac{\omega}{\kappa} \ln (\frac{\omega}{\kappa}) } \frac{|x|^{i\frac{\omega}{\kappa}}}{\sqrt{4\pi \omega}}.\label{minusout} 
\end{align}
On the right one gets, 
\begin{align}
\varphi_{\omega}^{\rm out} &\sim e^{i\frac{\omega}{\kappa} -i \frac{\omega}{\kappa} \ln (\frac{\omega}{\kappa}) } \frac{|x|^{i\frac{\omega}{\kappa}}}{\sqrt{4\pi \omega}},\label{plusout}\\
\varphi_{\omega}^{\downarrow} &\sim \frac{e^{\frac{\omega \pi}{2\kappa}}}{[2\Delta(x)]^{\frac12}} \frac{e^{-i\frac{\omega}{\kappa} \ln(\Lambda \sqrt{2\kappa})}}{\sqrt{4\pi \kappa(1- \kappa |x|)}} |x|^{-i\frac{\omega}{2\kappa}} e^{- \frac23 \Delta(x)}, \label{decmode} \\
\varphi_{\omega}^{\uparrow} &\sim \frac{e^{-\frac{\omega \pi}{2\kappa}}}{[2\Delta(x)]^{\frac12}} \frac{e^{-i\frac{\omega}{\kappa} \ln(\Lambda \sqrt{2\kappa})}}{\sqrt{4\pi \kappa(1- \kappa |x|)}} |x|^{-i\frac{\omega}{2\kappa}} e^{\frac23 \Delta(x)}.\label{grmode} 
\end{align}
We now comment these expressions. Firstly, the relative 
errors of the two {\it out} modes are
\be
O\left(\frac{\om^2/\kappa^2}{\Delta(x)^2} \right), \label{outerror}
\ee
whereas those of the four other modes are 
\be
O\left(\frac{1+\om/\kappa}{\Delta(x)} \right). \label{inerror}
\ee
To get these expressions,  we have taken into account two sources of errors: 
 those coming from the approximate roots of Sec.~\ref{approxHJ},
and those from the WKB approximation,  see App.~\ref{appWKB}. 
They all depend on  $\Delta$ of Eq. (\ref{Delta}). 
Again, we see that the errors on the {\it out} modes, of low momentum $p$, are subdominant with respect to those of
{\it in} modes which have a high momentum. As a result, $D_{\rm lin}^L$ of \Eqref{Dlin} will be more relevant than $D_{\rm lin}^R$. 

Secondly, whereas the normalization of the four oscillating 
modes is standard and based on 
the conserved scalar product of \Eqref{scalt}, those of 
the decaying $\varphi_{\omega}^{\downarrow}$ and growing $\varphi_{\omega}^{\uparrow}$ modes 
follow from \Eqref{xWKB} and the fact that $S_\om$ of \Eqref{Som} 
is complex since the roots of Eqs. (\ref{grroot}) and (\ref{decroot}) are. 

Thirdly, since these two modes do not appear in the `on-shell' Bogoliubov transformation of \Eqref{BogHR}, 
we should explain why we are still considering them. First, the forthcoming connection formula 
will be a transfer matrix relating the general solution on each side of the horizon.
Second, when considering problems with several horizons, these modes
could contribute to the `on-shell' S-matrix if they live in a finite (supersonic)  size region between two horizons~\cite{aQuattro}.

\subsection{Globally defined modes in the near horizon region}
\label{SecpWKB}
\subsubsection*{The $p$-representation}
To accurately describe the behavior of the modes across the horizon
one cannot used the $x$-WKB modes  
of the former Section.
Rather, one should work in $p$-space, and look for solutions of the form:
\be
\phi_{\omega}(x) = \int_{\mathcal C} \tilde{\phi}_{\omega}(p)\,  e^{ip x} \frac{dp}{\sqrt{2\pi}},
\label{ft}
\ee
where $\mathcal C$ is a contour in the complex $p$-plane. If it is well chosen, {\it i.e.} such that the integral converges and integrations by part can be performed, then it is sufficient 
that the dual mode $\tilde{\phi}_{\omega}$ satisfies 
\be
\left(- \omega + p \hat{v} \right) \left(- \omega + \hat{v}p \right)\tilde{\phi}_{\omega} = F^2(p) \tilde{\phi}_{\omega} \label{pmodequ},
\ee
where $\hat v = v(\hat x) = v(i\partial_p)$. We should notice that \Eqref{ft} is a standard Fourier transform only if 
$\mathcal C$ is on the real line, something we shall not impose. 
The main interest of considering generalized contours
is that it will enable us to compute {\it all} solutions of \Eqref{modequ}, including the growing ones.

Because we only need the behavior of the modes near the horizon, the operator
$\hat{v}$ in \Eqref{pmodequ} can  be replaced by $\hat v= -1 + i\kappa \partial_p$.
Hence one gets a second order ODE, irrespectively of $F(p)$.
The advantages to work in $p$-space are then clear~\cite{BMPS,Rivista05}. 
Firstly, the solutions of \Eqref{pmodequ} can be (exactly) written 
as a product 
\be
\tilde{\phi}_{\omega}(p) = \chi(p) \, e^{-i \frac{p}{\kappa}} \times 
\frac{p^{-i\frac{\omega}{\kappa} - 1}}{\sqrt{4\pi \kappa}},
\label{pchi}
\ee
where $\chi$ obeys the $\om$-independent equation:
\be
-\kappa^2 p^2 \partial_p^2 \chi = F^2(p) \chi(p),
\label{chiequ}
\ee
and where ${p^{-i\frac{\omega}{\kappa} - 1}}$ is independent of $F$. 
Hence, the deviations due to the dispersion $F$ are entirely encoded in $\chi$.
The origin of this factorization has to be found in the 
underlying structure of de Sitter space, see footnote \ref{f1}.
In addition, when considering the limit $\Lambda \to \infty$ in \Eqref{modequ}, 
${p^{-i\frac{\omega}{\kappa} - 1}}$ is exactly the relativistic (conformaly invariant) mode in $p$ space. 

Secondly, unlike the original equation in $x$, \Eqref{chiequ} is perfectly regular. Moreover, 
when dispersion effects are weak, {\it i.e.} $\Lambda/\kappa \gg 1$, the $p$-WKB approximation,
so called not to confuse it with that used in the former Section, is very good.
It generalizes what is done for the Airy function~\cite{AbramoSteg,Olver} 
where the mode equation in $p$-space is WKB exact. 
At this point, it is worth comparing the expression of the $p$-WKB modes in general and near the horizon. 
Using \Eqref{Wom}, one finds 
\ba
\tilde{\varphi}_{\omega}(p) &\sim& \sqrt{\frac{\partial X_\omega(p)}{\partial \omega}} \frac{e^{iW_\om(p)}}
{\sqrt{4\pi  F(p) }}, \label{YYY} 
\\
&=& \frac{e^{i \left(- \frac{\omega}{\kappa} \ln(p) + \int \frac{F(p') - p'}{\kappa p'}dp' \right)}}{\sqrt{4\pi \kappa\, 
p F(p)}} \times \left(1+O\left( \frac{\kappa}{\Lambda}\right) \right) \label{dualmode}.
\ea
The first line generalizes the standard WKB expression,
and can be obtained by Fourier transforming \Eqref{xWKB} at the saddle point approximation.
It shows that $\tilde{\varphi}_{\omega}$ is universally governed by 
$W_\om$ and $X_\om$, irrespectively of $F(p)$ and $v(x)$. 
The second line shows how exactly $F(p)$ 
enters in $\tilde{\varphi}_{\omega}$ in the near horizon region. In this region, since the mode equation is second order in $\partial_p$, 
the corrections are uniformly {\it bounded} by $\kappa/\Lambda$~\cite{Olver}. 
We also notice that using $\chi^*$ instead of $\chi$ in \Eqref{pchi} would describe a left moving mode 
of negative norm~\cite{Rivista05}. This shows that the corrections to the 
$p$-WKB approximation describe creation of pairs of left and right moving modes, as in cosmology~\cite{MacherCosmo}. 
It is also of interest to notice that in models where left and right movers stays completely decoupled~\cite{BMPS,SU08}, 
the $p$-WKB modes of \Eqref{dualmode} are {\it exact} solutions in the near horizon region. 

We finally notice that 
\Eqref{dualmode}, as the relativistic mode ${p^{-i\frac{\omega}{\kappa} - 1}}$, 
is well defined only when having chosen the 
branch cut of $\ln p$~\cite{BMPS,Corley,Tanaka99,SU}. As explained below, different possibilities, and different contours ${\cal C}$, lead to different modes. 

\section{Connection formula}
\label{CF}
\subsection{The various modes in the near horizon region} 
Using $F(p)$ of \Eqref{dispr} and \Eqref{dualmode}
one gets
\be
\varphi_{\omega}(x) = \frac1{\sqrt{4\pi \kappa}} \int_{\mathcal C} \frac{e^{i (px- \frac{\omega}{\kappa} \ln(p) + \frac{p^3}{6\Lambda^2 \kappa})}}{(1+\frac{p^2}{2\Lambda^2})^{\frac12}} \frac{dp}{p\sqrt{2\pi}} \label{contourmode}.
\ee
The forthcoming analysis generalizes former treatments for several  
reasons: 
\begin{enumerate}
\item Unlike~\cite{Corley,SU,Tanaka99}, we shall consider contours that are not homotope to the real line.
This will allow us to obtain the general connection formula which includes the growing mode.
\item We will make use of mathematical theorem of asymptotic development~\cite{Olver} 
 under their exact form. This will lead to the proper identification of the validity conditions in Sec.\ref{validity}.
\item We will compute the {\it phases} of the Bogoliubov coefficients. 
These are essential to compute the correlation pattern (see \Eqref{correl}) and 
in the presence of several horizons, see App.\ref{BHL}.
\end{enumerate}

To evaluate \Eqref{contourmode}, the first thing to take care of is the convergence of the integral. Indeed, for large $p$, the dominant term is of the shape $e^{ip^3}$, hence, we should impose that our contour $\mathcal C$ goes to infinity in regions where $\Im(p^3) > 0$. The second step is to perform the integral at a saddle point approximation.  For this, we make a change of variable such that the $p$ and $p^3$ terms are of the same order for all values of 
$\Lambda/\kappa$. We thus write $p = \Lambda \sqrt{2 \kappa |x|}t$ and get:
\be
\varphi_{\omega}(x) = \frac{e^{-i\frac{\omega}{\kappa} \ln(\Lambda \sqrt{2\kappa |x|})}}{\sqrt{4\pi \kappa}} \int_{\mathcal C} \frac{e^{-i\frac{\omega}{\kappa} \ln(t)}}{(1+t^2 \kappa |x|)^{\frac12}} e^{i\Delta(x) \left({\rm sign}(x) t + \frac{t^3}3\right)} \frac{dt}{t\sqrt{2\pi}}.
\label{t}
\ee
Hence we see that the large parameter $\Delta(x)$ defined in \Eqref{Delta} can be used to perform a saddle point approximation (see $z$ in \Eqref{SP}). Therefore $\Delta$ will govern the deviations due to this approximation.

For completeness we recall the saddle point theorem. If $z$ is some large parameter:
\be
\int_{\mathcal C} A(p) e^{i z f(p)} \frac{dp}{\sqrt{2\pi}} = \sum_j A(p_j) \frac{e^{i z f(p_j)}}{\sqrt{-if''(p_j) z}} \ \ \left(1+ O\left( \frac{E_j(f,A)}z \right)\right) \label{SP},
\ee
where the $p_j$ are saddle points of $f$, {\it i.e.} $f'(p_j) = 0$ of smallest imaginary part, and 
where the square root takes its principal value. 
This formula is valid if and only if one can deform $\mathcal C$ such that $\Im \left( f(p) - f(p_j) \right)$ is always 
positive~\cite{Olver}. The first correction $E_j(f,A)$ involves higher derivatives of $f$ and $A$ evaluated at $p_j$ 
\be
E_j(f,A) = \left(-i \frac{A''}{f''} + i\frac{f'''\, A'}{(f'')^2} -i \frac{5(f''')^2A - 3f''''\, f''\, A}{12(f'')^3} \right).
\ee

\subsubsection{The decaying mode}
\label{decayingSec}
We saw in Sec.\ref{trajcla} that for negative frequency, the particle is reflected near the horizon, at the turning point of \Eqref{Xp}.
Hence we expect that the corresponding mode will decay on the other side, for $x>0$.
This behavior is implemented by imposing that the branch cut is $-i\mathbb R_+$ and that the contour is homotope to the real line, 
as we now show.

To evaluate \Eqref{contourmode} for  $x>0$, we use \Eqref{t} and perform a saddle point calculation. The saddle points obey $1+t^2 = 0$. Just like in the Airy case~\cite{AbramoSteg,Olver}, 
only $t = i$ is relevant and its contribution is 
\be
\varphi_{\omega}(x) = e^{-i\frac{\pi}2} \frac{e^{\frac{\omega \pi}{2\kappa}}}{[2\Delta(x)]^{\frac12}} \frac{e^{-i\frac{\omega}{\kappa} \ln(\Lambda \sqrt{2\kappa}) }}{\sqrt{4\pi \kappa(1- \kappa |x|)}} |x|^{-i\frac{\omega}{2\kappa}} e^{- \frac23 \Delta(x)} \times \left( 1 + O\left(\frac{1+\frac{\omega^2}{\kappa^2}}{\Delta(x)} \right) \right). \label{firstSP}
\ee
As required, the mode decays on this side of the horizon. The error term has been estimated using
\Eqref{SP}. (More precisely, when computing $E_j$, we found a bounded function of $\omega/\kappa$ and $x$ times 
$(1+ \omega^2/\kappa^2)/\Delta$. This justifies our expression.) 
We notice that this expression coincides to $\varphi_{\omega}^{\downarrow}$ 
of \Eqref{decmode} up to a 
factor. Therefore, this factor defines the scattering coefficient 
and its $x$-dependent correction:  
\be
\varphi_{\omega} = \varphi_{\omega}^{\downarrow} \times \left( e^{-i\frac{\pi}2} \right)\times 
\left( 1 + O\left(\frac{1+\frac{\omega^2}{\kappa^2}}{\Delta(x)} \right) \right). 
\ee
We underline that this identification introduces no new errors because those due to the saddle point
are of the same order as those already present in \Eqref{decmode}. 
Since the general method is now understood, we proceed with the same mode on the other side of the horizon, and then apply the same method for two other modes
so as to get the general connexion formula.

\begin{figure}[!h]
\begin{center} 
\includegraphics[scale=0.55]{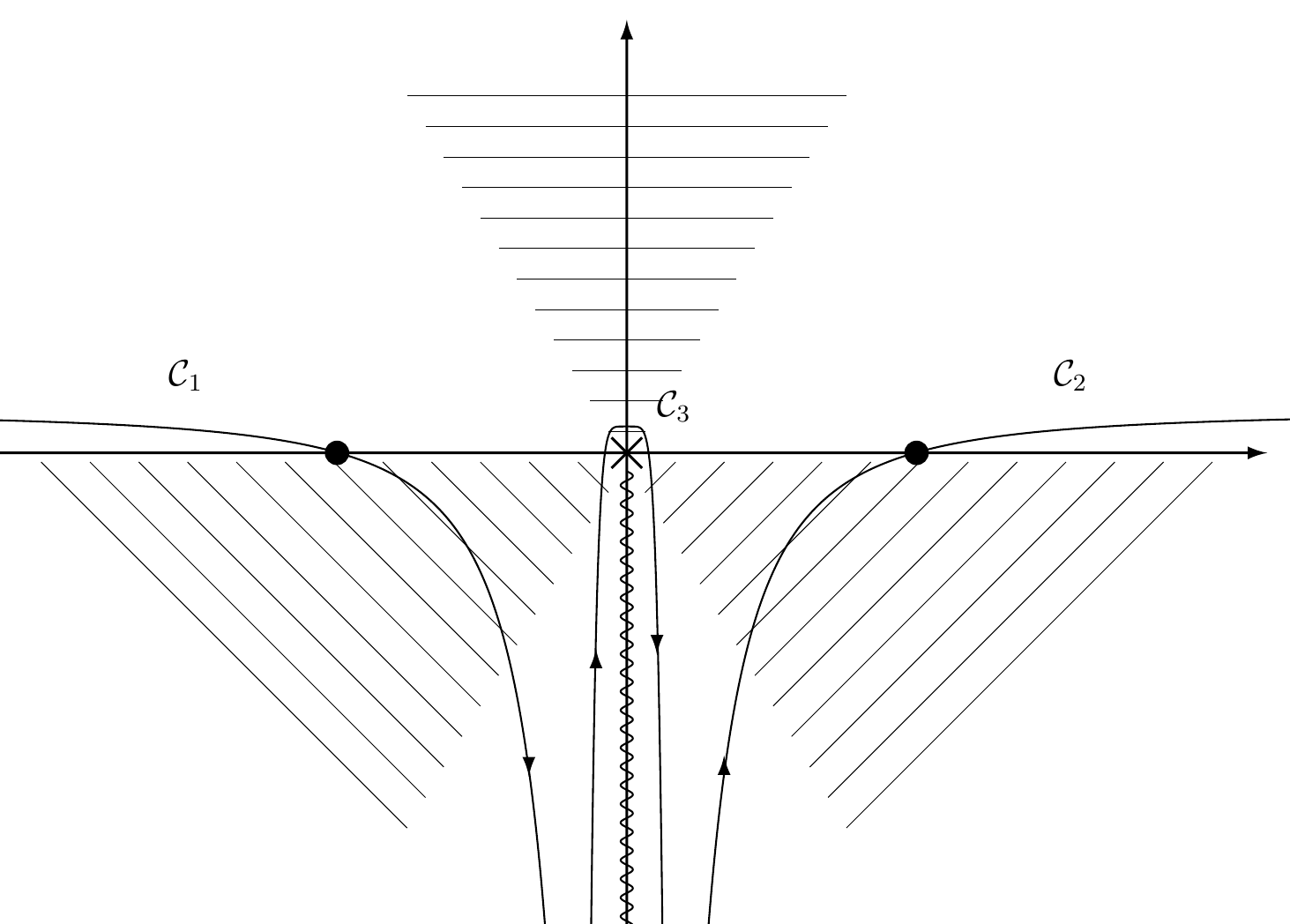}
\includegraphics[scale=0.55]{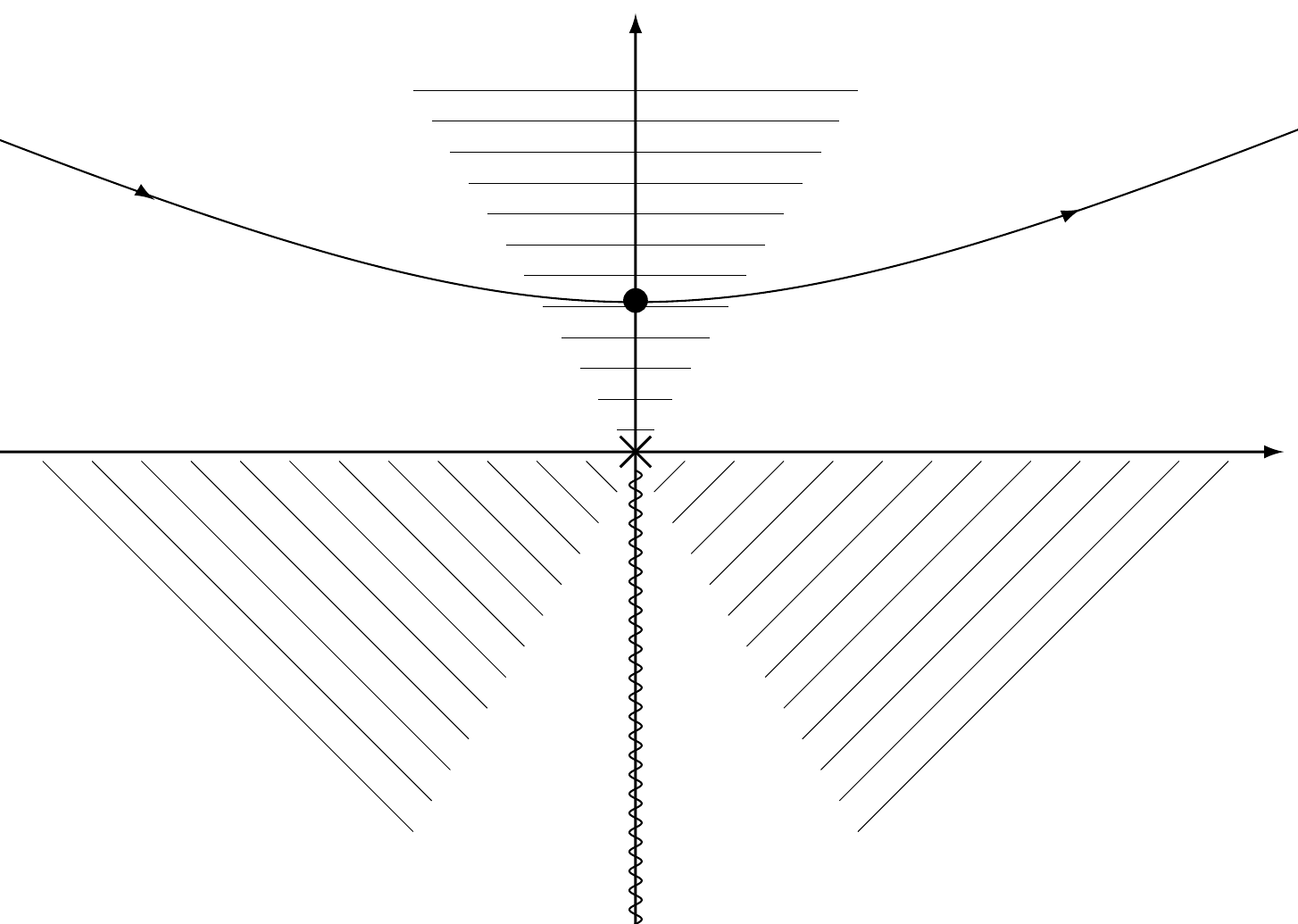}
\end{center}
\caption{Representation of the contours in the complex $p$-plane determining the decaying mode, 
on the left pannel, when $x$ is negative, and on the right when it is positive. 
The hatched regions are the asymptotically forbidden ones, and the dots indicate the saddle points
that contribute to the integral. }
\label{contours} 
\end{figure}

For $x < 0$, the saddle points now verify $1-t^2 = 0$. However, because of the branch cut, one must 
cut the contour into three separate branches, as shown in Fig.\ref{contours}. ${\mathcal C_1}$ and ${\mathcal C_2}$
 go from $\pm \infty + \epsilon$ and dive toward $-i\infty$ on each side of the branch cut. 
${\mathcal C_3}$ encircles the branch cut, and is necessary for the union of the $3$ new contours to be homotope to the original one. Separating the three contributions, 
$\varphi_{\omega} = \varphi^{\mathcal C_1} + \varphi^{\mathcal C_2} + \varphi^{\mathcal C_3}$,
$ \varphi^{\mathcal C_1}$ and  $\varphi^{\mathcal C_2}$ are evaluated by the saddle point method and, after identification with the WKB modes of \Eqref{plusin} and  \Eqref{minusin}, respectively give, 
\begin{align}
\varphi^{\mathcal C_1} &=\left(\varphi_{-\omega}^{\rm in}\right)^* \times ( e^{\frac{\omega \pi}{\kappa}} e^{i\frac{3\pi}4} )\times \left( 1 + O\left(\frac{1+\frac{\omega^2}{\kappa^2}}{\Delta(x)} \right) \right)
\label{C1},\\
\varphi^{\mathcal C_2} &=  \varphi_\omega^{\rm in} \times e^{i\frac{\pi}4}
\times  \left( 1 + O\left(\frac{1+\frac{\omega^2}{\kappa^2}}{\Delta(x)} \right) \right).\label{C2}
\end{align}
To properly evaluate $\varphi^{\mathcal C_3}$, one cannot use the saddle point method. 
However, because the factor $e^{ipx}$ decays along 
$\mathcal C_3$, one can use a `dominated convergence theorem', 
 {\it i.e.} take the limit $\Lambda \rightarrow \infty$ in the integrand of \Eqref{contourmode}. 
Using the Euler function, and $\varphi_{-\omega}^{\rm out}$ of \Eqref{minusout} we get
\begin{align}
\varphi^{\mathcal C_3} =& \left( \varphi_{-\omega}^{\rm out} \right)^* \times
 \left(- \sinh \left(\frac{\omega \pi}{\kappa} \right) \sqrt{\frac{2\omega}{\pi \kappa}} 
\Gamma\left( -i \frac{\omega}{\kappa}\right)e^{\frac{\omega \pi}{2\kappa}} e^{-i\frac{\omega}{\kappa} +i\frac{\omega}{\kappa} \ln(\frac{\omega}{\kappa})} \right)
\nonumber \\
&\times \left( 1 + O\left( \frac{\kappa |x|(1+\om^3/\kappa^3)}{\Delta(x)^2} \right) \right). \label{gammalike}
\end{align}
The correction term has been calculated by expanding the integrand in \Eqref{contourmode} to first order in $\kappa/\Lambda$ and computing the integral again with the use of the $\Gamma$ function.

\subsubsection{The transmitted mode}
To get another mode, we keep the same contour but 
the branch cut is now taken to be $i\mathbb R_+$. 
As we shall see it corresponds to a transmitted mode. For $x< 0$ the saddle points still obey $1-t^2 = 0$, but we can now use the saddle point approximation 
because the branch cut is no longer in the way. 
Taking into account that on the negative real $t$-axis $\ln t = \ln|t| -i\pi$, we get: 
\be
\varphi_{\omega} = \left[  \varphi_\omega^{\rm in}\times e^{i\frac{\pi}4}
+ \left( \varphi_{-\omega}^{\rm in}\right)^* 
\times  e^{-\frac{\omega \pi}{\kappa}} e^{i\frac{3\pi}4} 
\right] \times \left( 1 + O\left(\frac{1+\frac{\omega^2}{\kappa^2}}{\Delta(x)} \right) \right).
\ee
On the other side, for $x>0$, the saddle points obey $1+t^2 = 0$, as previously. 
Because of the branch cut one cannot pick the contribution of the decaying saddle point. 
We must instead deform the contour to a region where the `dominated convergence theorem' can be used and then stick it to the branch cut. 
With a computation similar to what was done for \Eqref{gammalike}, we find 
\begin{align}
\varphi_{\omega} &= \varphi_{\omega}^{\rm out}
\times \left(\sinh \left(\frac{\omega \pi}{\kappa}\right) \sqrt{\frac{2\omega}{\pi \kappa}}
\Gamma \left( -i \frac{\omega}{\kappa}\right) e^{-\frac{\omega \pi}{2\kappa}} e^{-i\frac{\omega}{\kappa} +i\frac{\omega}{\kappa} \ln(\frac{\omega}{\kappa}) } \right) \nonumber \\
&\times \left( 1 + O\left( \frac{\kappa |x|(1+\om^3/\kappa^3)}{\Delta(x)^2}\right) \right).
\end{align}

\subsubsection{The growing mode}
To get a third linearly independent solution of \Eqref{modequ}, we must construct the growing mode.
To get it, we re-use the above defined contours $\mathcal C_1$ and $\mathcal C_2$,
and we choose the branch cut to be $i\mathbb R_-$. 
For $x<0$, the relevant saddle points are $-1$ for $\mathcal C_1$ and $+1$ for $\mathcal C_2$, 
and respectively give \Eqref{C1} and \Eqref{C2}. 
For $x>0$ instead, for both contours, the relevant saddle point is $t=-i$, and this gives 
\begin{align}
\varphi^{\mathcal C_2}_{\omega} = - e^{- \frac{2\om \pi}{\kappa}} \, \varphi^{\mathcal C_1}_{\omega} = 
\varphi_\omega^{\uparrow} \times \left( 1 + O\left(\frac{1+\frac{\omega^2}{\kappa^2}}{\Delta(x)} \right) \right) .
\end{align}
Since all combinations of $\varphi^{\mathcal C_1}_{\omega}$ and $\varphi^{\mathcal C_2}_{\omega}$
behave as $ \varphi_\omega^{\uparrow}$  when $\Delta \to \infty$,
there is an ambiguity in choosing the third 
mode we shall use. We appeal to the conserved Wronskian to fix the choice. 
For a forth order differential equation, the Wronskian is a $4\times 4$ determinant, but because we neglected the $v$-modes, it becomes $3\times 3$. 
Once we have chosen 2 propagating modes, there is a {\it unique} choice of growing mode such that the basis has 
a unit Wronskian. 
The connection matrix is then 
an element of the group $SL_3(\mathbb C)$, {\it i.e.} of unit determinant (up to an overall gauge phase).

Therefore, our third mode is 
$\varphi_{\omega}^g= \varphi_{\omega}^{\mathcal C_2} - 
e^{-\frac{2\om \pi}{\kappa}} \varphi_{\omega}^{\mathcal C_1}$. 
On the right and on the left we respectively get: 
\begin{align}
\varphi_{\omega}^g &\underset{x>0}{=} 2 \varphi_{\omega}^{\uparrow}\times \left( 1 + O\left(\frac{1+\frac{\omega^2}{\kappa^2}}{\Delta(x)} \right) \right),\nonumber \\
& \underset{x<0}{=} \left[ e^{i\frac{\pi}4} \varphi_{\omega}^{\rm in} + e^{-\frac{\omega \pi}{\kappa}} e^{-i\frac{\pi}4} \left( \varphi_{-\omega}^{\rm in} \right)^* \right] \times \left( 1 + O\left(\frac{1+\frac{\omega^2}{\kappa^2}}{\Delta(x)} \right) \right).\label{Bilike}
\end{align}

\subsection{Connection matrix and on-shell Bogoliubov transformation} 
\subsubsection{The connection formula} 
The results of the former subsection can be synthesized 
by the following $3 \times 3$ `off-shell transfer matrix' that connects  
the WKB modes defined on either side of the horizon
\be
\begin{pmatrix}\varphi_{\omega}^{\rm out} \\ \varphi_{\omega}^{\downarrow} \\ \varphi_{\omega}^{\uparrow} \end{pmatrix} = (U_{\rm BH})^T \cdot \begin{pmatrix}\varphi_{\omega}^{\rm in}  \\ \left( \varphi_{-\omega}^{\rm in} \right)^* \\ \left( \varphi_{-\omega}^{\rm out} \right)^* \end{pmatrix} \label{Udef}.
\ee
We define $U_{\rm BH}$ through its transpose so that it 
relates the  
three amplitudes of {\it any} mode decomposed on the left and right side basis of Sec.\ref{basis}. 
Ignoring for the moment the correction terms, the matrix is
\be
U_{\rm BH} = \begin{pmatrix}  \tilde{\Gamma} \left(\frac{\omega}{\kappa} \right)^{-1} & e^{i\frac{3\pi}4} & \frac{e^{i\frac{\pi}4}}2 \\ e^{-\frac{\omega \pi}{\kappa}} e^{i\frac{\pi}2} \tilde{\Gamma} \left(\frac{\omega}{\kappa} \right)^{-1} & -e^{i\frac{\pi}4} e^{\frac{\omega \pi}{\kappa}}  & e^{-\frac{\omega \pi}{\kappa}} \frac{e^{-i\frac{\pi}4}}2 \\ 0 & e^{\frac{\omega \pi}{\kappa}} e^{-i\frac{\pi}4} \tilde{\Gamma} \left(\frac{\omega}{\kappa} \right) & 0  \end{pmatrix}.
\label{U}\ee
To simplify the above, we defined the `normalized' $\Gamma$ function:
\be
\tilde{\Gamma}(z) =  \Gamma \left( -iz \right)\, \sqrt{\frac{2z}{\pi}} \sinh(\pi z) e^{-\frac{\pi z}2} e^{i z \ln(z) - i z} e^{-i \frac{\pi}4},
\label{Gr0}
\ee
which obeys for large $z$
\begin{align}
\left| \tilde{\Gamma}(z) \right|^2 &\ = 1 - e^{-2\pi z},\\
{\rm Arg}\left(\tilde{\Gamma}(z) \right) &= 0 + \frac{1}{12z} + O \left(\frac{1}{z^2}\right).
\label{Gr}
\end{align}
As expected from our choice of modes, the determinant is a pure phase:
\be
\det (U_{\rm BH}) = e^{i \frac{\pi}2}.
\ee

\subsubsection{Robustness of black hole radiation}
Using \Eqref{U} we can now easily extract the Bogoliubov coefficients of \Eqref{BogHR}. Let us start with $\phi_\om^{\rm in}$. Being a physical mode, it is 
asymptotically bounded, and therefore the 
amplitude multiplying the growing mode should vanish. Moreover, on the left it asymptotes to 
$\varphi_\om^{\rm in}$ of \Eqref{plusin}. Hence its six amplitudes obey
\be
\begin{pmatrix} 1 \\ 0 \\ \beta_\om \end{pmatrix}_{x<0} = U_{\rm BH} \cdot \begin{pmatrix} \alpha_\om \\ d_\omega \\ 0 \end{pmatrix}_{x>0},
\ee
where $d_\om$ is the amplitude of the decaying mode. 
From these equations, and the corresponding ones for 
the negative frequency mode $\left(\phi_{-\omega}^{\rm in} \right)^*$,
the coefficients of \Eqref{BogHR} are
\begin{align}
\alpha_\omega &= \frac{\tilde{\Gamma} \left(\frac{\om}{\kappa}\right)}{1 - e^{-\frac{2\om \pi}{\kappa}}}= 
e^{ -i\frac{\pi}2}\, \tilde{\alpha}_\omega \nonumber ,\\
{ \beta_\omega \over  \alpha_\om } &=  e^{-\frac{\om \pi}{\kappa}} = 
{\tilde \beta_\omega \over \tilde \alpha_\om } . \label{bogocoef2}
\end{align}
The amplitudes of the decaying modes in $\phi_\om^{\rm in}$ and
 $\left(\phi_{-\omega}^{\rm in} \right)^*$ are also fixed and given by 
 \begin{align}
d_\om &= e^{i\frac{\pi}4} \frac{e^{-\frac{\om \pi}{\kappa}}}{2 \sinh(\frac{\om \pi}{\kappa})},
\nonumber \\
d_{-\om} &= e^{i\frac{3\pi}4} \frac1{2 \sinh(\frac{\om \pi}{\kappa})}.
\end{align}

For $\om \ll \om_{\rm max}$ of \Eqref{ommax},
and when ignoring the $x$-dependent corrections of the former subsection
({\it i.e.} to leading order in $\kappa/\Lambda$),
the mean occupation number \Eqref{e13} 
is exactly the relativistic result of \Eqref{relats}. 
This is in agreement with what was found in~\cite{BMPS,Corley,Tanaka99,SU}, although the conditions are now stated more precisely.
This result implies that the spectral deviations due to dispersion
are to be found by examining the various approximations that have been used. 

In addition, it should be noticed that sufficiently far away from the horizon, 
{\it i.e.} in a region where $\Delta(x)$ of \Eqref{Delta} is larger than 1,
the space-time correlation pattern of the Hawking particles of positive frequency and their inside
partners of negative frequency is also unaffected by dispersion. 
This second aspect of the robustness of HR can be established
by either forming wave packets of {\it in}-modes~\cite{BMPS}, {\it i.e.} 
considering non-vacuum states described by coherent states, see App.C. in~\cite{MacherBEC},
or by computing the 2-point correlation function 
$G(t,x; \, t'x')= \langle \phi(t,x)\, \phi(t',x') \rangle$~\cite{Carusotto2008,Balbinot2008,UnruhCor}.
For a comparison of the two approaches, see~\cite{From2010}. 
In the {\it in}-vacuum, at equal time, the $\om$ contribution of $G$ 
for $x> 0$ and $x' < 0$
is given by
\be
G_\om(x,x') = \alpha_\om \beta_\om^* \, 
\varphi_\om^{\rm out}(x) \varphi_{-\om}^{\rm out}(x') + \tilde \beta_\om \tilde \alpha_\om^* \left( \varphi_\om^{\rm out}(x)\right)^* \left( \varphi_{-\om}^{\rm out}(x')\right)^* \label{correl},
\ee
see the ${\cal B}_\om$ term in Sec.IV.F. in~\cite{MacherBEC}. 
Using the expressions of Sec.\ref{basis} together with \Eqref{bogocoef2}, far away from the turning point of \Eqref{xtp} but still in the near horizon region, we get 
\be
G_\om(x,x') = {1 \over \sinh(\frac{\om \pi}{\kappa})} \frac{\Re |x/x'|^{i\frac{\om}{\kappa}}}{4\pi \om} \times \left(1+ O \left(\frac{1+\om^2/\kappa^2}{\Delta(x)}\right)\right).
\ee
At leading order in $\kappa/\Lambda$, this is exactly the relativistic result. Hence, 
the long distance correlations are also robust when 
introducing short distance dispersion. This follows from the fact that 
the {\it phase} of $\beta_\om/\alpha_\om$ (and not just its norm) is not modified by dispersion. 
At this point we should say that this phase actually depends 
on those of the {\it in} and {\it out} modes that can be arbitrarily chosen. 
Thus, as such it is not an observable quantity. However, the phase of $\alpha_\om \beta_\om^* \varphi_\om^{\rm out} \varphi_{-\om}^{\rm out}$ is an observable,  
independent of these choices. We have chosen to work with {\it in} and {\it out} bases 
where all modes have a common phase at a given $p$, see Sec.\ref{actionSec},
as this ensures that $\arg(\beta_\om/\alpha_\om)$ is unaffected by this arbitrary phase. 

Using these basis, the phases of $\alpha_\om$ and $\tilde \alpha_\om$ of \Eqref{bogocoef2} also have a clear meaning as can be seen by 
considering the scattering  of classical waves. They characterize the phase shifts which are not taken 
into account by the WKB modes of Sec.~\ref{basis}.
In fact, using \Eqref{Gr0} one verifies that in the limit $\om/\kappa \to \infty$ 
one recovers the standard WKB results, {\it i.e.} $\arg(\alpha_\om) =\arg(\tilde\Gamma(\om/\kappa)) \to 0$ 
for the transmitted mode, and
$\arg(\tilde \alpha_\om)  \to \pi/2$ for the reflected one. 
For smaller values of $\om/\kappa$, 
$\arg(\tilde \Gamma(\om/\kappa))$ thus accounts for the non-trivial phase shift.
In App.~\ref{BHL} we show that in the presence of two horizons, this shift affects the
spectrum of trapped modes.

\subsubsection{White holes}
\label{WHSec}

It is rather easy to apply our results 
to white holes as they behave as the time reverse of black holes. 
In terms of stationary modes, the correspondence is made 
by exchanging the role of {\it in} and {\it out} and perform a complex conjugation:
\be
\phi_\om^{\rm in, WH} = \left( \phi_\om^{\rm out, BH}\right)^*,
\ee
as in Eq.~(D3) of~\cite{MacherBEC}. This follows from a symmetry of \Eqref{modequ}. 
Indeed, the mode equation is invariant under
\be
\om \to -\om \text{ and } v \to -v \label{BHWHsym}.
\ee
For more details, we refer to the App.D of~\cite{MacherBEC}. 
Here, it implies 
\be
U_{\rm WH} = \left(U_{\rm BH}\right)^* \label{BHtoWH},
\ee
where $U_{\rm WH}$ is defined through the same equation as \Eqref{Udef}.
This implies that Bogoliubov coefficients of a white hole 
posses the same norm as those in the corresponding black hole setup.


It is more interesting to look at correlations. Indeed, since both 
{\it out} modes now live on the same side of the horizon and contain high momenta, 
the correlation pattern drastically differs from the black hole one~\cite{Mayoral2011}. 
To describe this pattern, in order not to introduce confusing notations, the modes 
shall be called according to their status in a black hole geometry, 
that is, in the following expressions the modes are those of Sec.\ref{basis}. Then,  
in the white hole {\it in}-vacuum, the $\om$-contribution of the equal time correlation  
function reads 
\be
G^{\rm WH}_\om(x,x') = \phi_\om^{\rm out}(x) \left(\phi_\om^{\rm out}(x')\right)^* + \phi_{-\om}^{\rm out}(x) \left(\phi_{-\om}^{\rm out}(x')\right)^* \label{Gom}.
\ee
Decomposing the BH {\it out} modes in their {\it in} content, the first term gives
\begin{align}
\phi_\om^{\rm out}(x) \left(\phi_\om^{\rm out}(x')\right)^* =& |\alpha_\om|^2 \phi_\om^{\rm in}(x) \left(\phi_\om^{\rm in}(x')\right)^* - \alpha_\om^* \tilde \beta_\om \phi_\om^{\rm in}(x) \phi_{-\om}^{\rm in}(x'),
\nonumber \\ 
&- \alpha_\om \tilde \beta_\om^* \left(\phi_{-\om}^{\rm in}(x)\right)^*\left( \phi_\om^{\rm in}(x')\right)^* + |\tilde \beta_\om|^2 \left(\phi_{-\om}^{\rm in}(x)\right)^* \phi_{-\om}^{\rm in}(x').
\end{align}
The most interesting phenomenon occurs when one looks at the zero frequency limit. 
In this regime, on the one hand, the zero frequency limit of $\phi_{\om}^{\rm in}$ and $\phi_{-\om}^{\rm in}$ 
agree with each other, as can be seen in \Eqref{plusin} and \Eqref{minusin}, 
and equal $\phi_0^{\rm in} \in \mathbb C$ which is a non-trivial function of $x$. On the other hand, when $\om \to 0$, \Eqref{bogocoef2} 
gives 
\be
|\alpha_\om|^2 \sim |\tilde \beta_\om|^2 \sim \frac{\kappa}{2\pi \om} \ 
\text{ and }\ - \alpha_\om^* \tilde \beta_\om \sim e^{-i\frac\pi2}\frac{\kappa}{2\pi \om}. 
\ee
Then, since both terms of \Eqref{Gom}
agree in the limit, we finally get
\be
G^{\rm WH}_\om(x,x') \underset{\om \to 0}{\sim} \frac{4\kappa}{\pi \om} 
\Re\left\{ e^{-i\frac\pi4}\, \phi_0^{\rm in}(x)\right\} \, \Re\left\{ e^{-i\frac\pi4} \, 
\phi_0^{\rm in}(x') \right\},  
\label{Gwh}
\ee
Before commenting on this expression, it is interesting to compare it with 
the corresponding one obtained in a black hole geometry. In that case, when 
working in the black hole {\it in} vacuum, the $\om$ contribution of the two point function is, see \Eqref{Gom},
\be
G^{\rm BH}_\om = \phi^{\rm in}_\om(x)\left( \phi^{\rm in}_\om(x')\right)^* + \phi^{\rm in}_{-\om}(x) \left(\phi^{\rm in}_{-\om}(x')\right)^* .
\ee 
Hence in the limit $\om \to 0$ it gives (see Eq.~32 in~\cite{UnruhCor})  
\be
G^{\rm BH}_0 = 2\phi^{\rm in}_0(x) \left( \phi^{\rm in}_0(x')\right)^*,
\ee
which behaves very differently than $G^{\rm WH}$ of \Eqref{Gwh}. It does not diverge as $1/\om$
and it is not the product of two {\it real} waves. 
To get an expression that might correspond to that of \Eqref{Gwh}
one should express the {\it in} modes in terms of the {\it out} black hole modes given in \Eqref{minusout} and (\ref{plusout}), as done in \Eqref{correl}. 
Doing so, the prefactor of \Eqref{Gwh} is recovered but the spatial behavior of $G^{\rm BH}$
is completely different because the {\it out} modes of \Eqref{minusout} and (\ref{plusout})
are defined on opposite sides of the horizon,
and become constant in their domain. Hence unlike $G^{\rm WH}$,
$G^{\rm BH}$ cannot be written in the limit $\om \to 0$ as a product of twice the same real wave.

In \Eqref{Gwh}, we see that 
$G^{\rm WH}$ factorizes in the zero frequency limit 
as the 2-point function in inflationary cosmology when neglecting the {\it decaying} mode, 
see {\it e.g.} Eq.~(20) in~\cite{CampoPar}. 
This fact shows that $\Re\left\{e^{-i\frac\pi4}\,  \phi_0^{\rm in}(x)\right\}$
contributes to $G^{\rm WH}$ as 
in
 a stochastic ensemble of {\it classical} waves, 
{\it i.e.} each member of the ensemble contributes 
like a coherent state with a given phase, see App.C in~\cite{MacherBEC},
but the amplitude of $\phi_0^{\rm in}$ is still a 
Gaussian random variable, again as for primordial fluctuations
in cosmology. Moreover, the real function $\Re\left\{e^{-i\frac\pi4}\,  \phi_0^{\rm in}(x)\right\}$ corresponds to
the {\it undulation} observed in~\cite{SilkePRL2010}, and the present analysis
shows that this phenomenon is directly related to the Hawking effect. 
However, what fixes the constant amplitude found in the experiment remains to be understood 
because, when summing \Eqref{Gwh} over $\om$, the factor $1/\om$ engenders  
a logarithmic growth in $\ln(t)$ as in Ref.~\cite{Mayoral2011}. 
\Eqref{Gwh} differs from the expression of~\cite{Mayoral2011} (which also factorizes) 
in that the surface gravity has not been sent to $\infty$. 
Using Sec.\ref{basis}, one can compute its profile (in the WKB approximation) 
 in the regions 1.(b) of Fig.\ref{regions}. In the right region 1.(b) it decays according to the zero frequency limit of \Eqref{firstSP},  
whereas in the left region, one has
\be
\Re\left\{e^{-i\frac\pi4}\, \varphi_0^{\rm in}(x)\right\} = \frac1{\sqrt{8\pi \kappa}} \frac{\cos(\frac23\Delta(x)+\frac\pi4)} {\sqrt{\Delta(x)(1+\kappa |x|)}} .
\label{undulpi}
\ee
On the right it thus 
behaves very much like the decaying Airy function $Ai$~\cite{Olver,AbramoSteg}, 
and on the left it oscillates in a similar manner but, quite surprisingly, 
the phase shift $\pi/4$ has the opposite sign.
The origin of this flip is to be found in the extra factor of $1/p$ 
in the integrand of \Eqref{contourmode} that is associated with the relativistic (non-positive) 
norm of \Eqref{scalt}. It would be very interesting to observe 
the profile of \Eqref{undulpi} and its unusual phase shift in future experiments.
Further away from the horizon, the undulation profile 
can be obtained from the zero frequency limit of \Eqref{xWKB}.

\subsection{Validity of the connection formula}
\label{validity}
Our computation is based on two approximations. The first one is the $p$-WKB approximation introduced 
when solving \Eqref{chiequ} in the near horizon region. Its validity requires
\be
\frac{\Lambda}{\kappa} \gg 1.
\ee
This condition is the expected one. 
It involves neither $\om$ nor the parameters $D_{\rm lin}$ of \Eqref{Dlin}.
Moreover, as we shall see below, 
it will {\it not} be the most relevant one in the general case. This is a non-trivial result. 
In addition, since the corrections to this approximation 
mix left and right moving modes~\cite{Rivista05},
at leading order, the same spectral deviations will be 
obtained when considering models~\cite{BMPS,SU08} where the decoupling between these modes is exact. 
It would be interesting to validate this prediction by numerical analysis.

The second approximations are controlled by 
$\Delta$ of \Eqref{Delta}. This quantity governs both the validity of the saddle point approximation, 
as in \Eqref{firstSP}, and that of the WKB  
modes of Eq. (\ref{plusin}-\ref{grmode}).
Since these corrections decrease when $\Delta$ increases, 
the pasting of the near horizon modes on the 
WKB ones should be done at the edges of the near horizon region. 
One could imagine pasting the modes further away, but this would require to control 
$\tilde \phi_\om(p)$ outside the region where $v$ is linear in $x$, {\it i.e.} to deal 
with ODE in $\partial_p$ of order higher than $2$. This is perhaps possible
but it requires other techniques than those we used, see~\cite{scottthesis} for recent developments.  
In any case, what we do is sufficient to control the error terms in the relativistic limit, and more precisely to find an upper bound.

\begin{figure}[!h]
\begin{center} 
\includegraphics[scale=2]{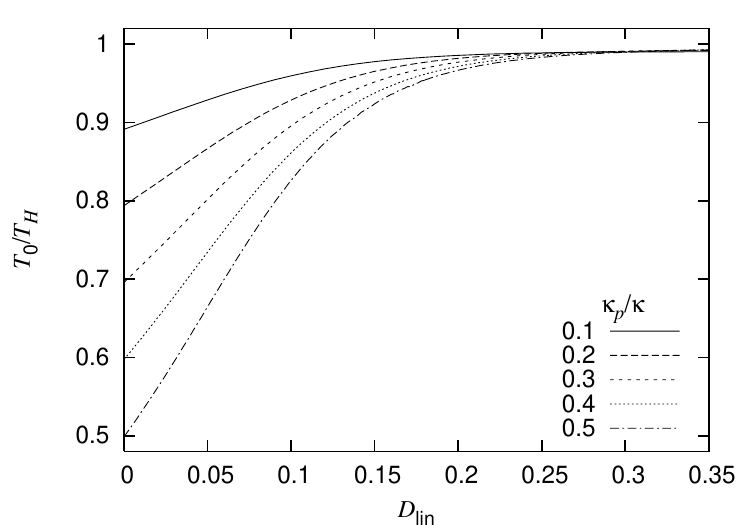}
\end{center}
\caption{Deviation of the temperature found using 
the code of~\cite{FP2} for various flows $v(x)$ which all
have the same surface gravity $\kappa$.
$T_H = \kappa/2\pi$ is the usual Hawking temperature. 
$T_0$ is the actual temperature given quartic superluminal dispersion
computed for $\om \ll \kappa$.
The parameter $\kappa_p$ (see~\cite{FP2} for its precise definition) 
characterizes the slope of $v(x)$ {\it outside} the near horizon region, {\it i.e.} in regions 2 of Fig.\ref{regions}. 
At fixed $\Lambda/\kappa=15$, the critical value of $D_{\rm lin} \sim 0.2$ 
below which $T_0$ deviates from $T_H$ does not depend on $\kappa_p$. This is in agreement with \Eqref{Dp}. }
\label{Xlinfig} 
\end{figure}

Being confined to stay within the near horizon region, the validity of the pasting procedure requires that 
\be
\Delta_{\rm p} \equiv
\Delta(x_{\rm pasting}) =\frac{\Lambda}{\kappa} (\kappa x_{\rm pasting})^{\frac32} 
 \sim \frac{\Lambda}{\kappa}\left(D_{\rm lin}\right)^{\frac32} \gg 1,
\label{Dp}
\ee
where $D_{\rm lin}$ characterizes the extension of this region on either side of the horizon.
As we see in Fig.\ref{Xlinfig}, at fixed $\Lambda/\kappa$, the spectrum is very close to the relativistic one of \Eqref{relats} if 
 $D_{\rm lin}$ is large enough. Instead, below a certain threshold, the deviations become non negligible. Our criterion in \Eqref{Dp}  
indicates that this will happen when $\Delta_{\rm p}$ is of order 1. 
Hence the threshold value for $D_{\rm lin}$ should scale as $({\kappa}/{\Lambda})^{2/3}$. This prediction 
is confirmed by the numerical analysis of~\cite{FP2}.

We can be more precise. 
Indeed, as discussed after \Eqref{vlin},  
the near horizon region is not necessarily symmetric. Hence, the values of $x_{\rm pasting}$ 
on the right and on the left will be different. 
This is important because error terms coming from {\it in} modes 
are dominant compared to those from {\it out} modes,
see Sec.\ref{xmodeanalys}. Since the former lives on the left side, 
the validity condition is
\be
\Delta_{\rm p}^L \sim \frac{\Lambda}{\kappa}\left(D_{\rm lin}^L\right)^{\frac32} \gg 1,
\label{DpL}
\ee
The higher sensitivity of the spectrum to perturbations of $v$ localized on the left hand side 
was clearly observed in~\cite{FP2}, see Fig. 8 right panel. This sensitivity has been
recently exploited in~\cite{Zapata2011} 
to produce resonant effects. We can now estimate the deviations on the spectrum. 
Considering Eqs. (\ref{outerror},\ref{inerror},\ref{firstSP},\ref{gammalike},\ref{DpL}), 
we obtain
\be
\left|\frac{\bar n_\om - \bar n_\om^{\rm relativistic}}{\bar n_\om^{\rm relativistic}}\right| = O\left( \frac{P(\om/\kappa)}{\Delta_p^L} \right)\label{leadcorr}, 
\ee
where $P$ is a polynomial function of degree 2. Therefore, at fixed $\om \lesssim \kappa$, {\it i.e.} in the relevant regime for Hawking radiation, 
the leading deviations are {\it bounded} by a quantity scaling as $1/\Delta_{\rm p}^L$. 
In particular, \Eqref{leadcorr} demonstrates that the spectrum is thermal for arbitrary small frequencies,
contrary to what has been claimed in~\cite{River,Corley}. 
Moreover, our analysis indicates that the deviations 
should grow with $\om$. This is compatible with the fact that $\bar n$ vanishes for 
$\om > \om_{\rm max}$ defined in \Eqref{ommax}, irrespectively of the value of the ratio $\Lambda/\kappa$.
However, we have not yet been able to confirm this precisely with the code of~\cite{FP2}. 
One of the reason is that the spectrum is radically modified when $\om $ approaches $\om_{\rm max}$. 
So far the $p$-WKB approximation has been a subdominant effect. However, 
if one takes the de Sitter limit, {\it i.e.} $D_{\rm lin} \to \infty$, the correction term in \Eqref{Dp} 
vanishes and $p$-WKB becomes the only source of deviations.

To conclude we recall that the deviations have been computed using \Eqref{modequ}.
In other mode equations, like that for phonons in a BEC~\cite{MacherBEC},
the corrections to the $x$-WKB approximation will be in general larger. 
However these corrections are not due to dispersion but rather
to the fact that the conformal invariance of  \Eqref{modequ} in the dispersionless limit
will be lost. These corrections can thus be studied without introducing dispersion.
This is also true when introducing an infrared modification of \Eqref{modequ} associated with a 
mass or with a non-vanishing perpendicular momentum. 
In App.\ref{massApp}, we show that the spectral properties are still robust in that \Eqref{DpL} is sufficient to guarantee 
that to leading order the Bogoliubov coefficients are unaffected by short distance dispersion. 

\subsection{General superluminal dispersion}
\label{generalization}
Instead of \Eqref{dispr}, we now consider
\be
F^2(p) = \left(p + \frac{p^{2n+1}}{\Lambda^{2n}} \right)^2 \label{generalF},
\ee
taken again, for simplicity, to be a perfect square. 
In Sec.\ref{SecpWKB} we computed p-WKB modes for any dispersion in \Eqref{dualmode}, hence the globally defined 
modes for $F$ satisfying \Eqref{generalF} are, see Sec.\ref{CF}, 
\be
\phi_{\omega}(x) = \frac1{\sqrt{4\pi \kappa}} \int_{\mathcal C} \frac{e^{i (px- \frac{\omega}{\kappa} \ln(p) + \frac{p^{2n+1}}{(2n+1)\Lambda^{2n} \kappa})}}{(1+\frac{p^{2n}}{\Lambda^{2n}})^{\frac12}} 
\frac{dp}{p\sqrt{2\pi}}.
\ee
There now exist $2n+1$ linearly independent modes. 
Hence in terms of contours, there are $2n+1$ sectors (Stokes lines) 
toward $\infty$ in the complex $p$ plane. 
By using a contour homotope to the real line, and the same two possible branch cuts of $\ln p$
on $\pm i\mathbb R^+$, we can compute the $2$ `on-shell' modes 
that are asymptotically bounded. Even though, there  
exist $n$ pairs of growing and decaying modes on the subsonic side, and $n-1$ 
pairs on the other side, only one pair 
in the subsonic sector will be relevant in the `off-shell' connection formula of \Eqref{U}. 
Indeed, all the others pairs do not mix with propagating modes. 
Therefore, the different contours giving rise to the relevant modes 
will be quite similar to those of Sec.\ref{CF}. 

To perform a saddle point approximation, we introduce $t= p/ \Lambda|\kappa x|^{\frac{1}{2n}}$ 
and get:
\be
\varphi_{\omega}(x) = \frac{e^{-i\frac{\omega}{\kappa} \ln(\Lambda|\kappa x|^{\frac{1}{2n}})}}{\sqrt{4\pi \kappa}} \int_{\mathcal C} \frac{e^{-i\frac{\omega}{\kappa} \ln(t)}}{(1+t^{2n} \kappa |x|)^{\frac12}} e^{i\Delta_n(x) \left({\rm sign}(x) t + \frac{t^{2n+1}}{2n+1}\right)} \frac{dt}{t\sqrt{2\pi}} \label{modifdual}.
\ee
By a computation similar to that of Sec.\ref{CF}, 
at leading order in $\kappa/\Lambda$,
we recover the Bogoliubov coefficients of \Eqref{bogocoef2}, thereby establishing their robustness for arbitrary integer values of $n$.
Moreover, the deviations from this result are now governed by
\be
\Delta_{p, n}^L = \frac{\Lambda}{\kappa} \left( D^L_{\rm lin}
\right)^{\frac{2n+1}{2n}}.
\ee
Hence, the error on the mean number of emitted quanta satisfies 
\be
\left|\frac{\bar n_\om - \bar n_\om^{\rm relativistic}}{\bar n_\om^{\rm relativistic}}\right| = O\left( \frac{\kappa}{\Lambda (D_{\rm lin}^L)^{\frac{2n+1}{2n}}}\right) \label{corrgeneral}.
\ee

\subsection{Relating subluminal dispersion relations to superluminal ones}

So far we analyzed only superluminal dispersion relations. We should thus inquire 
how would subluminal dispersions affect the spectrum. At the classical level, as noticed at the end of Sec.~\ref{trajcla}, there is a exact correspondence between these two cases. 
At the level of the modes \Eqref{subsuper} does not leave \Eqref{modequ} invariant 
as it does not apply to the left moving solutions
governed by $\om - vp = - F$. However, it becomes a symmetry 
when neglecting the mode mixing 
between left and right movers. Therefore, in models where the decoupling between these is exact~\cite{BMPS,SU08}, 
\Eqref{subsuper} is an {\it exact} symmetry. 
Moreover, since  the mode mixing between left and right movers is subdominant  
for general mode equations, the discrepancy of the spectral deviations between superluminal and subluminal dispersion 
will not show up at leading order. This is {\it precisely} what has been observed in Sec.VI.2 of~\cite{Macher1}.

At the level of the connection formula of \Eqref{U}, the three exchanges of \Eqref{subsuper}
still are an exact symmetry since the $U_{\rm BH}$-matrix 
is based on the right moving mode of \Eqref{dualmode}
which is determined by the action $W_\om(p)$. 
Therefore,  at leading  order in $\kappa/\Lambda$, without any further calculation, 
this symmetry implies that the spectrum of HR is equally robust for subluminal dispersion.
Moreover, the {\it leading order deviations} from the thermal spectrum will be 
governed by the {\it same} expression as \Eqref{corrgeneral}. 
It should be noticed that when applying \Eqref{subsuper}, 
$D_{\rm lin}^L$ characterizes the extension on the sub-sonic region. 

The above symmetry should not be confused with that of \Eqref{BHWHsym} which also relates black and white holes. 
Indeed, the latter exchanges the role of left and right moving modes, while the former applies only to the right moving sector. 
Instead, these two symmetries can be {\it composed} with each other. This allows to compare black hole spectra 
without referring to white holes. 

To conclude the discussion, we mention that this approximate symmetry allows 
not only to {\it predict} several effects, but also to predict 
how the observables will quantitatively behave:
 \begin{itemize}
\item
a laser effect will be found for sub-luminal dispersion in a flow possessing two horizons 
that passes from super to sub and then back to a supersonic 
flow,\footnote{We are grateful to Daniele Faccio and William Unruh
for bringing our attention to this possibility in a discussion that took place 
during the Nice Colloquium on `Analog gravity' in June 2010.} and this 
exactly for the same reasons that the laser effect was found in the `reversed' flow in 
the case of superluminal dispersion~\cite{BHlaser} and App.\ref{BHL}.
\item 
the frequencies and the growth rates of this subsonic laser effect will be governed by the 
same expressions as those of App.\ref{BHL} 
(when neglecting the coupling to the left moving modes).
\item
subsonic phonon propagation in a non-homogeneous flow that remains everywhere subsonic,
{\it i.e.} without a sonic horizon, will be governed by a $4\times 4$ S-matrix that encodes new pair creation
channels with respect to those found in the presence of a sonic horizon~\cite{scottthesis} 
exactly for the same reasons that a supersonic phonon propagation in a non-homogeneous flow that remains everywhere supersonic does so, as mentioned in~\cite{FP3}. 
\item
the behavior of the Bogoliubov coefficients in the two cases will behave {\it quantitatively} 
in the same way.
\item
the undulation observed in white hole flow for subluminal gravity waves 
in the experiment of~\cite{SilkePRL2010} is generated for the 
same reasons as that found in white holes for Bose condensates using the Bogoliubov-de~Gennes 
equation~\cite{Mayoral2011}. As we discussed above, this is obtained by composing the symmetries of \Eqref{BHWHsym} and \Eqref{subsuper}.
\item
Moreover, when the dispersion relations and the profiles $v(x)+c(x)$ 
(where $c(x)$ is the speed of sound that generalizes the $1$ in $1+v$) obey \Eqref{subsuper}
up to a possible rescaling of $\Lambda$, the momentum $p$, and distances, 
these undulations should have the same spatial profile.
\end{itemize}

\section{Conclusion}
In this paper, we described the scattering of a dispersive field on a stationary black hole horizon. 
We computed the connection formula which relates WKB modes on each side of the horizon sufficiently far away
from the turning point. Our main results are: 
\begin{itemize}

\item \Eqref{U} applies `off-shell', which means that the contribution of the growing mode is taken into account. When requiring that its amplitude vanishes, \Eqref{U} fixes both the `on-shell' 
$2\times 2$  matrix of \Eqref{BogHR} encoding the Bogoliubov transformation 
between physical modes, and the amplitude of the decaying mode. 
In situations with several horizons, the presence of a growing mode in a finite size region could alter the scattering. In that case, \Eqref{U} should be used to estimate this effect.

\item The phases of the scattering coefficients are computed. 
This allows us to show that not only the out-going flux is robust against dispersion, but also the correlation pattern between Hawking quanta and their partner (in \Eqref{bogocoef2} 
and \Eqref{correl}).
Moreover, in situations with several horizons, such as that of the black hole laser, these shifts enter
in interference effects and directly affect the observables. Our prediction for their values
is numerically confirmed as discussed in App.\ref{BHL}. 

\item We characterized the leading spectral deviations due to dispersion.
These critically depend on the spatial extension of the near horizon geometry,
and not only on the ratio $\kappa/\Lambda$. 
Indeed, they are governed by the parameter $D_{\rm lin}^L$ of \Eqref{Dlin}, and this for
for a large class of dispersion, see Sec.\ref{generalization}. These spectral deviations imply only an upper bound on the frequency, 
in contradiction to what was claimed in~\cite{River,Corley}.

\item 
A symmetry between superluminal and subluminal dispersion relations 
establishes that the spectral deviations due to one 
type of dispersion will be also found 
for the corresponding one when modifying the flow $v(x)$ in a certain manner, see \Eqref{subsuper}.

\item
In App.~\ref{massApp}, we extend our results to massive fields. In that case, the dispersion relation is modified both in the infrared and the ultraviolet. This induces two types of deviations from \Eqref{bogocoef2}. We show that these deviations decouple when $m\ll \Lambda$ and can be bounded separately. The ultraviolet effects are still controlled by \Eqref{Dp} while the infrared ones can be studied with the massive relativistic equation. 

\end{itemize}

\subsection{Acknowledgments}
We acknowledge interesting discussions with Ulf Leonhard, Stefano Liberati and Scott Robertson. We also thank 
Roberto Balbinot, Iacopo Carusotto and Alessandro Fabbri for interesting discussions. We are thankful to Germain Rousseaux for mentioning \cite{GRouss}. 
This work was partially supported by the ANR grant STR-COSMO, ANR-09-BLAN-0157. 

\appendix
\section{WKB dispersive modes}
\label{appWKB}
\subsection{The method}
To apply a WKB approximation to the solutions of \Eqref{modequ}, 
we write the mode as 
\be
\phi_\om(x) = e^{i\int^x k_\om(x') dx'}\, .
\label{wkbm}
\ee
Injecting this into \Eqref{modequ}, we obtain 
\ba
(\om - v(x) k_\om)^2 - F^2\left(k_\om \right) &=& -i \partial_x \left[\frac12 \partial_k F^2\left(k_\om\right) \right] \nonumber\\
& & - \frac{4 k_\om \partial_x^2  k_\om + 3 (\partial_x k_\om)^2 - i \partial_x^3 k_\om}{\Lambda^2} \label{rica}.
\ea
For definiteness and simplicity, the last term is given for 
$F(k) = k^2 + k^4/\Lambda^2$ but can be generalized to any polynomial dispersion relation. 
So far, this equation is exact. It is known as a Riccati equation \cite{Olver} 
and was already used in the present context in~\cite{Corley}. 
It is adapted to a perturbative resolution where
the different terms are sorted in order of derivatives~\cite{BirrelDavies,Gottfried}, here
spatial gradients. Hence we write $k_\om$ as
\be
k_\om = k_\om^{(0)} + k_\om^{(1)} + k_\om^{(2)} +...
\label{series}
\ee
where the superscript gives the number of derivatives (one way to 
sort the terms of \Eqref{rica} is to make the scale change $x \rightarrow \lambda x$. The superscript then stands for the power of $1/\lambda$).
It is easy to show that $k_\om^{(0)}(x)= p_\om(x)$, 
the classical momentum, solution of the Hamilton-Jacobi equation \eqref{disp2}.
The second equation is not less remarkable: 
it is a total derivative and it has the universal form
\be
k^{(1)}_\om = \frac i2 \partial_x \ln \left[\om v(x) - v^2p_\om +\frac12 \partial_p 
F^2(p_\om) \right] = \frac i2 \partial_x \ln \left[F(p_\om)\, v_{gr}(p_\om) \right],
\label{k1k0}
\ee
where $v_{gr} = 1/\partial_\om p_\om$ is the group velocity. 
Since $k_\om^{(1)}$ is purely imaginary, it governs the mode amplitude. 
\Eqref{wkbm} constructed with $k_\om = p_\om + k_\om^{(1)}$ 
 gives the generalized $x$-WKB expression 
\be
\varphi_\omega^{\rm WKB}(x) = \sqrt{\frac{\partial p_\omega}{\partial \omega}} \frac{e^{i\int^x p_\om(x') dx'}}{\sqrt{4\pi  F(p_\om) }}.
\label{xWKB}
\ee
Its Fourier transform evaluated at the saddle point gives the $p$-WKB mode of \Eqref{YYY}.

\Eqref{k1k0} guarantees 
that the scalar product of \Eqref{scalom} 
evaluated with \Eqref{xWKB} is {\it exactly} conserved.
\Eqref{rica} also guarantees that 
the development of $k$ is alternated: even terms 
are real, while odd ones are imaginary. 

\subsection{Leading deviations due to dispersion} 
In this paper, we are interested in solving \Eqref{modequ} in the limit of weak dispersion, 
that is, for $\Lambda$ large enough.  The aim of this section is to precise the meaning of 
`$\Lambda$ large enough' 
by computing the scaling of the errors made when building the WKB basis of Sec.\ref{basis}. 
Two different approximations have been used. 
First we built approximate solutions of \Eqref{modequ} with the above WKB modes. 
Second, we solved Eq.~(\ref{disp2}) in the limit of large $\Lambda$ in Sec.\ref{approxHJ},
and we used the approximate roots to compute the WKB modes. 

To proceed, we estimate the next order term of \Eqref{series} 
in the limit of weak dispersion (WD), {\it i.e.} by dropping terms of order 2 in $1/\Lambda$. 
In this regime, $k^{(2)}$ is solution of
\be
-2\left[ F(p_\om) \, v_{gr}(p_\om) \right] k_{WD}^{(2)}(x) = \left[ -i\partial_x + k_\om^{(1)} \right]\cdot \left[  k_\om^{(1)}\, \partial_k(F v_{gr}) \right] \label{k2}.
\ee
Because we master the modes near the horizon in the $p$-representation (see Sec.\ref{SecpWKB} and \ref{CF}), we are only interested in the error accumulated from infinity, where the $x$-WKB approximation becomes exact, to $x_{\rm pasting}$ of \Eqref{Dp}, at the edge of the near horizon region. 
This error is estimated by evaluating the integral of $k_{WD}^{(2)}$ from $x_{\rm pasting}$ till $\infty$. 
Indeed the exact mode $\phi_\om$ can be approximated by 
\be
\phi_\om \simeq \varphi^{\rm WKB}_\om e^{i\int^x k_{WD}^{(2)}(x') dx'} \simeq \varphi_\om^{\rm WKB}
(1 + \epsilon(x)). \label{WKBapprox}
\ee 
To evaluate $\varphi^{\rm WKB}_{\om}$ we use \Eqref{xWKB} and the approximate roots of Sec.\ref{approxHJ}. This introduces extra errors 
governed by $y$ of \Eqref{y}. Hence, near $x_{\rm pasting}$ the total error is 
\be
\epsilon_T  \simeq \int_{\infty}^{x_{\rm pasting}} k_{WD}^{(2)}(x') dx' + O(y(x_{\rm pasting})). 
\ee
This error term behaves quite differently for {\it in} and {\it out} modes, hence we shall study it 
separately. 
\begin{itemize}
\item For the {\it in} modes, solving the \Eqref{k2}, we get
\be
k_{WD}^{(2)} = \frac{9 v'^2}{16\Lambda |1+v|^{\frac52}} - \frac{3 v''}{4\Lambda |1+v|^{\frac32}}.
\ee
Since this is not a total derivative, 
the integral depends on what happens all along the way from $\infty$ to $x_{\rm pasting}$. However, when the profile $v$ is smooth enough, the accumulated error is essentially 
\be
\epsilon^{\rm in}_T \sim \left( \frac{v'}{\Lambda |1+v|^{\frac32}} \right)_{x = x_{\rm pasting}} +O(y_p) = \frac1{\Delta_p^L} + \frac{\om/\kappa}{\Delta_p^L}.
\ee

\item For the {\it out} modes, the leading order correction arises from $k^{(1)}$. 
Indeed, in the limit $\Lambda \to \infty$, the {\it out} modes are WKB exact because of conformal invariance 
and $k^{(1)}_{\Lambda \to \infty} = 0$. 
Therefore, for finite $\Lambda$, $k^{(1)}$ will be the dominant contribution to the error term of \Eqref{WKBapprox}. 
One finds
\be
k_{WD}^{(1)} = - \frac{6 \om^2 v'}{\Lambda^2 (1+v)^4}.
\ee
This means that
\be
\epsilon^{\rm out}_T \sim \left( \frac{\om^2}{\Lambda^2 |1+v|^3} \right)_{x = x_{\rm pasting}} = \frac{\om^2/\kappa^2}{\Delta_p^2}.
\ee
Here the pasting happens on both sides, hence in the latter expression, one should understand $\Delta_p^L$ for $\left(\varphi_{-\om}^{\rm out}\right)^*$ and $\Delta_p^R$ for $\varphi_\om^{\rm out}$. As expected, the corrections to red-shifted {\it out} modes are subdominant with respect to those of {\it in} modes. 

\end{itemize}

It is interesting to notice that  the validity of the $x$-WKB approximation in a given flow 
and for a given dispersion relation depends on the exact form of the mode equation associated with \Eqref{disp2}. 
Indeed, when the mode equation is not conformally invariant in the limit $\Lambda \to \infty$
there is an extra validity condition:
\be
\left| \frac{v'}{\om}\right| \ll 1.
\ee
It is due to the fact that for low frequencies, the left and right moving modes mix
even in the absence of dispersion. This mixing was studied in \cite{MacherBEC}, and it was numerically shown that 
these effects stay (in general) subdominant. 

\section{Application: The black hole laser}
\label{BHL}
\subsection{The setup}
When a flow contains both a black and a white horizon, and is subsonic on both asymptotic sides, 
HR is self-amplified in presence of superluminous dispersion~\cite{BHlaser,Ulflaser}, and 
this leads to a dynamical instability~\cite{lasermode}. 
When the horizons are well separated, a powerful `$S$-matrix' description 
can be applied to describe the mode propagation. 
In this picture, as in \Eqref{BogHR}, the matrix acts on the two 
right moving modes $\left( \varphi_\omega, \left(\varphi_{-\omega}\right)^*\right)$. 
Because the horizons are well separated, it is legitimate to factorize 
$S$ as $S=S_W S_2 S_B S_4$, where $S_W$ and $S_B$ 
are the `on-shell' Bogoliubov matrices associated with the white and black hole respectively, 
and $S_2$ and $S_4$ contain phase shifts characterizing Hamilton-Jacobi propagations 
from the white hole horizon to the black one. 
These have the form 
\begin{align}
S_2 &= \begin{pmatrix} e^{iS_\omega^{\rm in}} & 0 \\ 0 & e^{iS_{-\omega}^{\rm in}} \end{pmatrix} ,\\
S_4 &= \begin{pmatrix} 1 & 0 \\ 0 & e^{-iS_{-\omega}^{\rm out}} \end{pmatrix},
\end{align}
where $S_\omega^{\rm in}$ designates $S_\om$ of \Eqref{Som} evaluated for the root $k_\omega^{\rm in}$,
and similarly for the other two actions. The names of the roots are those
associated with the black hole horizon, so that their expressions can be found in Sec.\ref{approxHJ}.
We notice that the precise values of the end points of integration 
in these three actions depend on the phase conventions that have been adopted to compute  $S_W$ and $S_B$, 
so that $S$ is independent of these arbitrary choices. 
Notice also the minus sign before $S_4$ because the corresponding propagation is 
backwards in $x$. 
For a more detailed analysis, we refer the reader to~\cite{lasermode}. Here, we only recall the main points. 

For a real frequency $\om>0$, the eigen-mode is asymptotically a plane wave of positive norm taken to be unity. 
To compute $b_\om$, 
the amplitude of the negative norm mode trapped between the two horizons,
we require that the mode be {\it single valued}. 
This condition gives 
\be
\begin{pmatrix} e^{if_\omega} \\ b_\omega \end{pmatrix} = S \cdot \begin{pmatrix} 1 \\ b_\omega \end{pmatrix}.
\ee
where $e^{if_\om}$ is the transmission coefficient, here a phase shift.
In addition to the real spectrum, there 
exists a discrete set of pairs of complex frequency modes which encode the instability. 
To construct these modes, we look for 
frequencies $\lambda_a = \omega_a + i\Gamma_a $ (where $a$ labels a discrete set) 
such that the mode is asymptotically bounded (in fact square integrable). 
As demonstrated in~\cite{lasermode}, the unstable modes ($\Gamma >0$) are purely {\it out}-going and thus 
satisfy
\be
\begin{pmatrix} A_a \\ 1 \end{pmatrix} = S \cdot \begin{pmatrix} 0 \\ 1 \end{pmatrix}.
\ee
This equation has a solution if and only if
\be
S_{22}(\omega + i\Gamma) = 1. \label{S22}
\ee
This is the algebraic equation that fixes the set of complex frequencies $\la_a$. 

\subsection{Predictions}

To compute these complex frequencies, we perform an expansion 
in the $\beta_\om$ coefficients in $S_W$ and $S_B$. 
To zeroth-order in these coefficients, \Eqref{S22} gives a Bohr-Sommerfeld condition 
that fixes $\om_a$, the real part of the frequency:
\be
\int_{x_{tp}^{W}}^{x_{tp}^B} \left( p^{\rm out}_{-\omega}(x) - p^{\rm in}_{-\omega}(x)\right)dx 
- \arg(\tilde \alpha_\om^W \tilde \alpha_\om^B) = 2\pi n, \ n\in \mathbb N^* \label{BS},
\ee
where the end point $x_{tp}^W$ (respectively $x_{tp}^B$) 
refers to the turning point near the white hole (respectively black hole) horizon.~\footnote{The writing 
of \Eqref{BS} is obtained only if the Bogoliubov coefficients $\tilde \alpha_\om^W$ and $ \tilde \alpha_\om^B$
of \Eqref{BogHR}
are computed with a phase $\theta_0$ of \Eqref{theta0} common to {\it in} and {\it out} modes. It is 
now clear why it is convenient to work with this choice. }  
The l.h.s. of \Eqref{BS} displays both standard and unusual features. 
On the one hand, as expected, one finds the classical action evaluated along a closed loop 
from one turning point to the other one. One the other hand, one does not find the 
usual phase shift ($=\pi$) that accounts for the two reflections when 
dealing with Schrödinger type problems where the modes near the turning points 
can be well approximated by Airy functions. Indeed, using \Eqref{bogocoef2} and \Eqref{Gr0},
one sees that $-\arg(\tilde \alpha_\om^W \tilde \alpha_\om^B)$ differs from $\pi$.
To characterize the difference, we introduce
\begin{align}
\epsilon \left(\omega\right) &= \arg(\tilde \alpha_\om^W \tilde \alpha_\om^B) + \pi,\label{epsphase}\\
& = \arg(\tilde \Gamma({\om}/{\kappa_W})) + \arg(\tilde \Gamma({\om}/{\kappa_B})).
\nonumber 
\end{align}
Using \Eqref{Gr}, one sees that it is only in the 
limit $\om/\kappa \to \infty$ 
that one recovers the standard result, i.e. $\epsilon = 0$. For smaller values of
$\om/\kappa$, $\epsilon$ accounts for the non-trivial phase shift due to
the reflections on the two horizons. It arises from the fact that
the reflected modes cannot be well approximated by Airy functions, something not discussed in~\cite{GRouss}.

In Fig.~\ref{Gammalas} we have compared the numerical
results obtained using the code of~\cite{FP1} with the theoretical predictions evaluated with and without $\epsilon$. The improvement
of the agreement when including  $\epsilon$ is clear.
\begin{figure}[!h]
\begin{center} 
\includegraphics[scale=1]{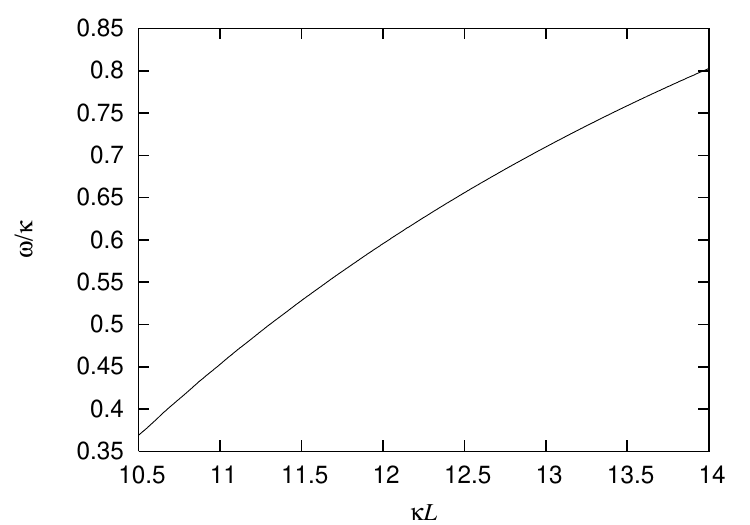}
\includegraphics[scale=1]{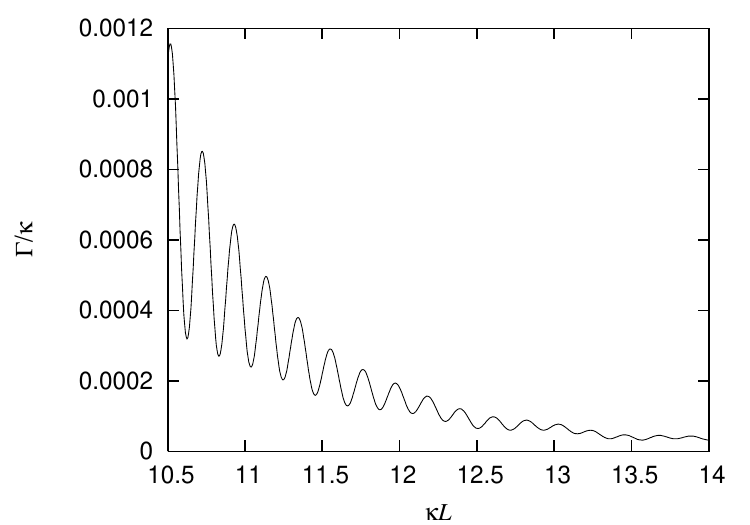}
\end{center}
\caption{Evolution of the real  part (left plot) and the imaginary (right plot) 
of a complex frequency as a function of $L$, the distance between the 2 horizons. These curves have been obtained by making used of the numerical techniques 
described in~\cite{FP1}. The parameters used are $\kappa_w/\kappa_b = 0.5$, $D=0.5$ and $\Lambda/\kappa_b = 8$, and we consider the $n=22$ discrete mode.}
\label{omegalas} 
\end{figure}
\begin{figure}[!h]
\begin{center}
\includegraphics[scale=1]{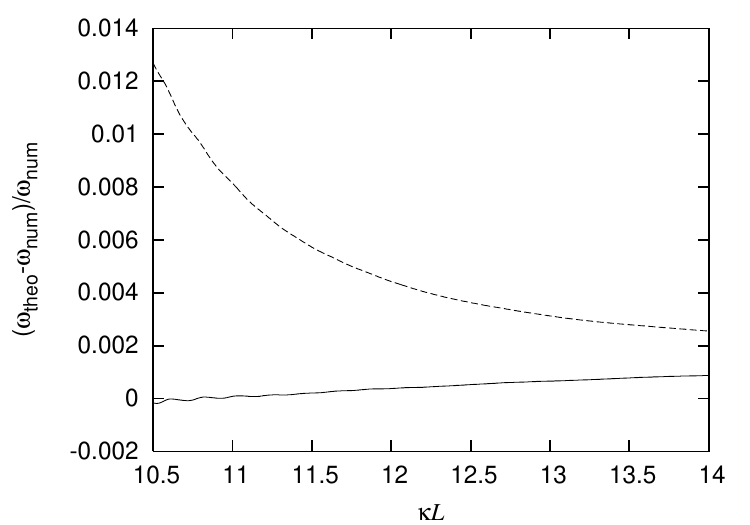}
\includegraphics[scale=1]{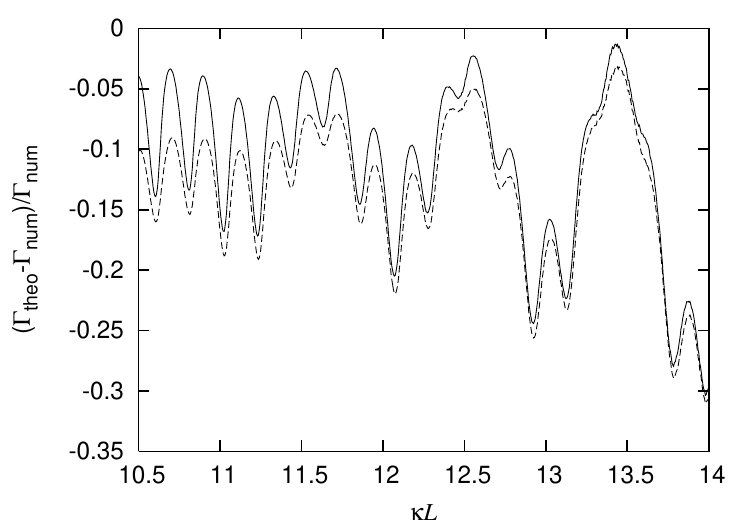}
\end{center}
\caption{Relative errors between the numerical results and our theoretical formulae
of the real and imaginary parts of the same complex frequency as in the above Figure.
The continuous lines take 
into account $\epsilon$ of \Eqref{epsphase}, while the dotted lines
are based on the standard expression $\epsilon = 0$. 
The improvement of the estimation is clear, and therefore 
the necessity of computing the phases in 
Eq.(\ref{bogocoef2}) is established.}
\label{Gammalas} 
\end{figure}

To second order in $\beta_\om$, 
\Eqref{S22} fixes the growing rate $\Gamma$ for a given value of $\om$:
\be
2 \Gamma T_\omega = \left| {\beta_\om^W \over \alpha_\om^W } \right|^2 +  \left| {\beta_\om^B \over \alpha_\om^B} \right|^2
+  2 \left| {\beta_\om^W \over \alpha_\om^W }{\beta_\om^B \over \alpha_\om^B}  \right|
\cos(\psi) \label{Gamma},
\ee
where $T_\om$ is the classical time for the trapped mode to make a round trip between the two horizons
\be
T_{\omega} = \partial_\omega \left( S^{\rm out}_{-\omega} - S^{\rm in}_{-\omega} + \pi - \epsilon(\om) \right),
\ee
and the phase $\psi$ reads
\be
\psi = S_\om^{\rm in} - S_{-\om}^{\rm in} + \arg \left( -\frac{\beta_\om^W \tilde \beta_\om^B }
{\alpha_\om^W \tilde \alpha_\om^B} \right) .
\ee
This non-trivial phase arises from the interference between the two pair creation amplitudes, that occurring at the white hole 
and that at the black hole.
A careful  calculation gives a remarkably simple result
\be
\psi = {\rm Re}\left\{ \int_{p_{\rm min}}^{p_{\rm max}} \left(X_\om^B(p) - X_\om^W(p) \right)dp \right\} 
+ \pi \label{psiphase}.
\ee
The constant phase shift follows from Eqs. (\ref{bogocoef2}, \ref{BHtoWH}) which give 
\be
\arg \left( -\frac{\beta_\om^W \tilde \beta_\om^B }{\alpha_\om^W \tilde \alpha_\om^B} \right) = \pi.
\label{pii}
\ee
The contribution from the classical actions is best expressed in 
$p$-space using the function $X_\om(p)$ of \Eqref{Xomp}. 
However, because the flow profile $v$ is not monotonic in the black hole laser setup,
$X_\om$ is no longer unique. Assuming that $v < 0$ has a single minimum 
at $x=0$ between the two horizons, 
we define $X_\om^W(p)$ (resp. $X_\om^B(p)$) as the solution of \Eqref{Xomp} 
of negative values describing the propagation towards the white hole 
(resp. positive values associated with the black hole). 
Both of these semi-classical trajectories run 
from a positive maximum value $p_{\rm max} = p_\om^{\rm in}(0)$ 
to a minimum negative value $p_{\rm min}  = p_{-\om}^{\rm in}(0)$. 
In \Eqref{psiphase} we took the real part of the integral in order to remove the imaginary
contributions ($= i\pi \om/\kappa_W$ and $i\pi \om/\kappa_B$)  that arise when $p$ flips sign, 
see the discussion after \Eqref{theta0}. 
The contribution for $p > 0$ accounts for the propagation of the positive 
norm mode, whereas that with $p < 0$ for that of the trapped mode.
What is remarkable is that when they are combined, the net result for $S_\om^{\rm in} - S_{-\om}^{\rm in}$ takes the 
form of the first term of \Eqref{psiphase}. Notice again that 
this simple form is found only when using mode bases such that \Eqref{pii} applies. 

Interestingly, \Eqref{psiphase}
 has the same structure as  Bohr-Sommerfeld condition 
of \Eqref{BS} with the role of $x$ and $p$ interchanged, {\it i.e.} \Eqref{psiphase} 
is a closed loop in $p$-space of some $X_\om(p)$. 
What is unusual is that the action receives imaginary contributions when $p$ changes sign.
In fact when removing the restriction to the real part of the action, one gets
\be
e^{i \int_{p_{\rm min}}^{p_{\rm max}} \left(X_\om^B(p) - X_\om^W(p) \right)dp + \pi}
= \frac{\beta_\om^W \tilde \beta_\om^B } {\alpha_\om^W \tilde \alpha_\om^B}\,  e^{i \psi }. 
\ee
In other words the ratios $\beta_\om/\alpha_\om$ can be blamed on, and therefore computed 
from, the imaginary contributions of the action $S_\om$ of \Eqref{Som} that arise when $p$ flips sign. 
An early version of this relation was used for relativistic modes in~\cite{DamourRuffini76},
and it is at the core of the so called `Unruh' modes~\cite{Unruh76}. 
It was adapted to dispersive waves in~\cite{BMPS} and implicitly used above 
when computing the connection 
coefficients in Sec.\ref{CF}. It was also recently exploited in~\cite{scottthesis} in a similar context.

To conclude, we underline that unlike the robustness of Hawking spectrum, 
\Eqref{BS} and \Eqref{psiphase} are
not necessarily valid for very small frequencies. Indeed, we have made two additional  
approximations. First that the $S$-matrix can be 
factorized, {\it i.e.} that the propagation is correctly described by separating the scattering on each 
horizon and by WKB propagations between the horizons. 
Second, that the growing rate of the unstable modes are small ($\Gamma T_\omega^b \ll 1$)
when developing Eq.~(\ref{S22}). 
This condition is violated when $\omega \rightarrow 0$ because $\beta/\alpha$ becomes of order 1. 
However, when the density of unstable modes is not too low, 
the numerical analysis confirms that the above expressions correctly
characterize the complex frequencies, as can be seen from 
Fig.\ref{omegalas} and \ref{Gammalas}.

\section{Massive case}
\label{massApp}

In this appendix we add a mass $m$ on top of the ultraviolet dispersion of \Eqref{dispr}:
\be
F_m^2  = m^2 + \left( p + \frac{p^3}{2\Lambda^2}\right)^2 ,
\ee
and focus on the novel features brought in by the mass. 
We suppose that the two scales are well separated, {\it i.e.} $ m \ll \Lambda$. 
In this limit one has 
\be
F_m(p) = \sqrt{m^2 + p^2} +\frac{p^4}{2\Lambda^2\sqrt{m^2 + p^2}}.
\ee
Moreover, since the last term becomes non negligible only when $p\sim \Lambda \gg m$, 
we get
\be
F_m(p) = \sqrt{m^2 + p^2} +\frac{p^3}{2\Lambda^2}.
\ee
Therefore, to first order in $m/\Lambda$, $F_m$ is a sum of the relativistic massive dispersion plus a dispersive term. 
In the near horizon region, the mode in $p$-space is still given by \Eqref{dualmode}, and, 
as in Sec. \ref{CF}, the various modes in $x$-space 
are given by contour integrals
\be
\varphi_{\omega,m}^{\mathcal C}(x) 
= \frac1{\sqrt{4\pi \kappa}} \int_{\mathcal C} \left(\frac{p}{F_m(p)}\right)^{\frac12} e^{i (px- \frac{\omega}{\kappa} \ln(p) + G_m(p) +\frac{p^3}{6\Lambda^2 \kappa} )} \frac{dp}{p\sqrt{2\pi}} \label{massmode},
\ee
where
\be
G_m(p) = -\int_p^{\infty} \frac{\sqrt{m^2+p'^2} - p'}{\kappa p'}dp',
\ee
encodes the modification of the phase due to the mass. 
As before, the choice of the contour $\mathcal C$ dictates which mode one is considering. 

\begin{figure}[!h]
\begin{center} \includegraphics[scale=0.5]{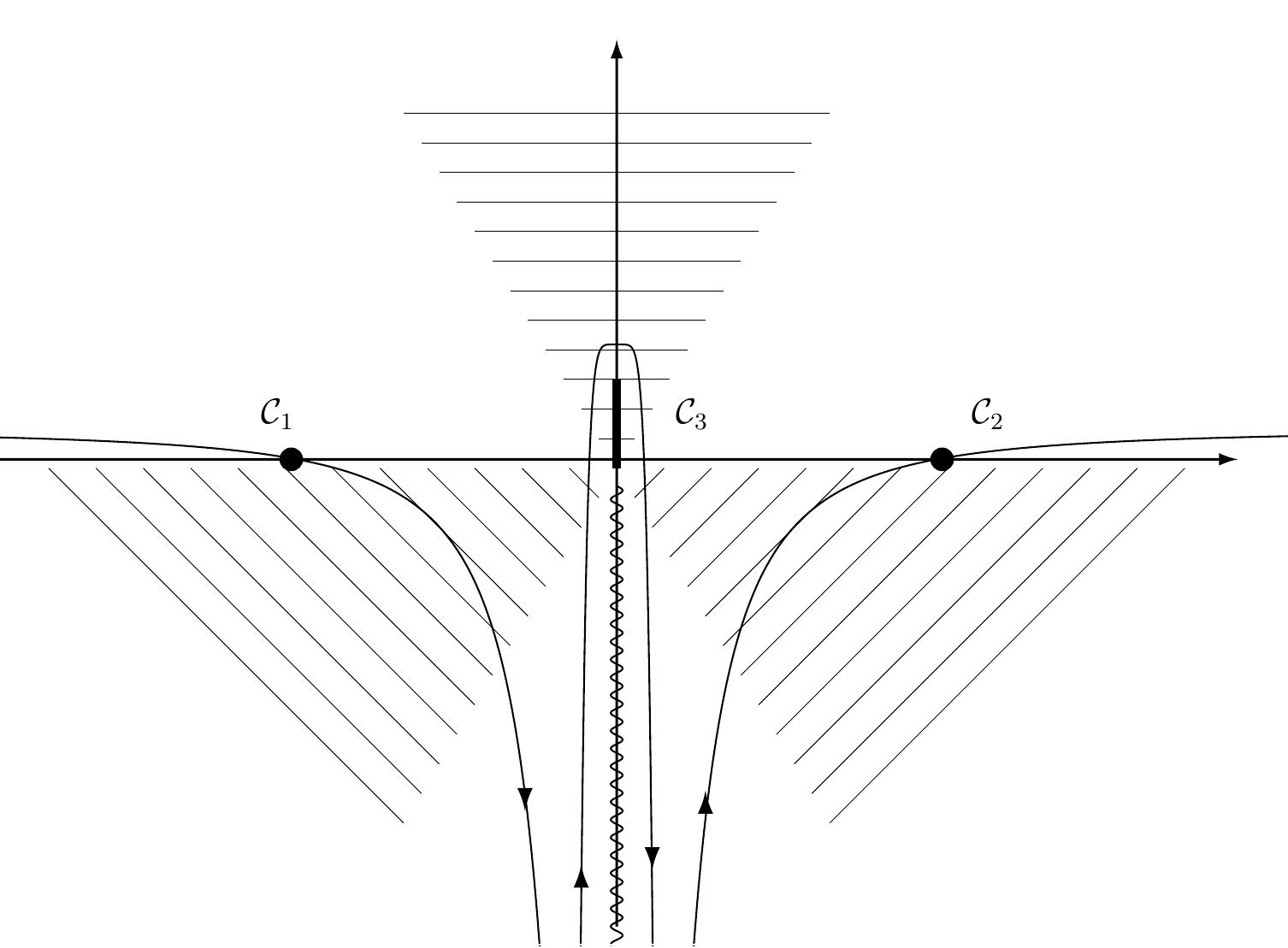}
\includegraphics[scale=0.5]{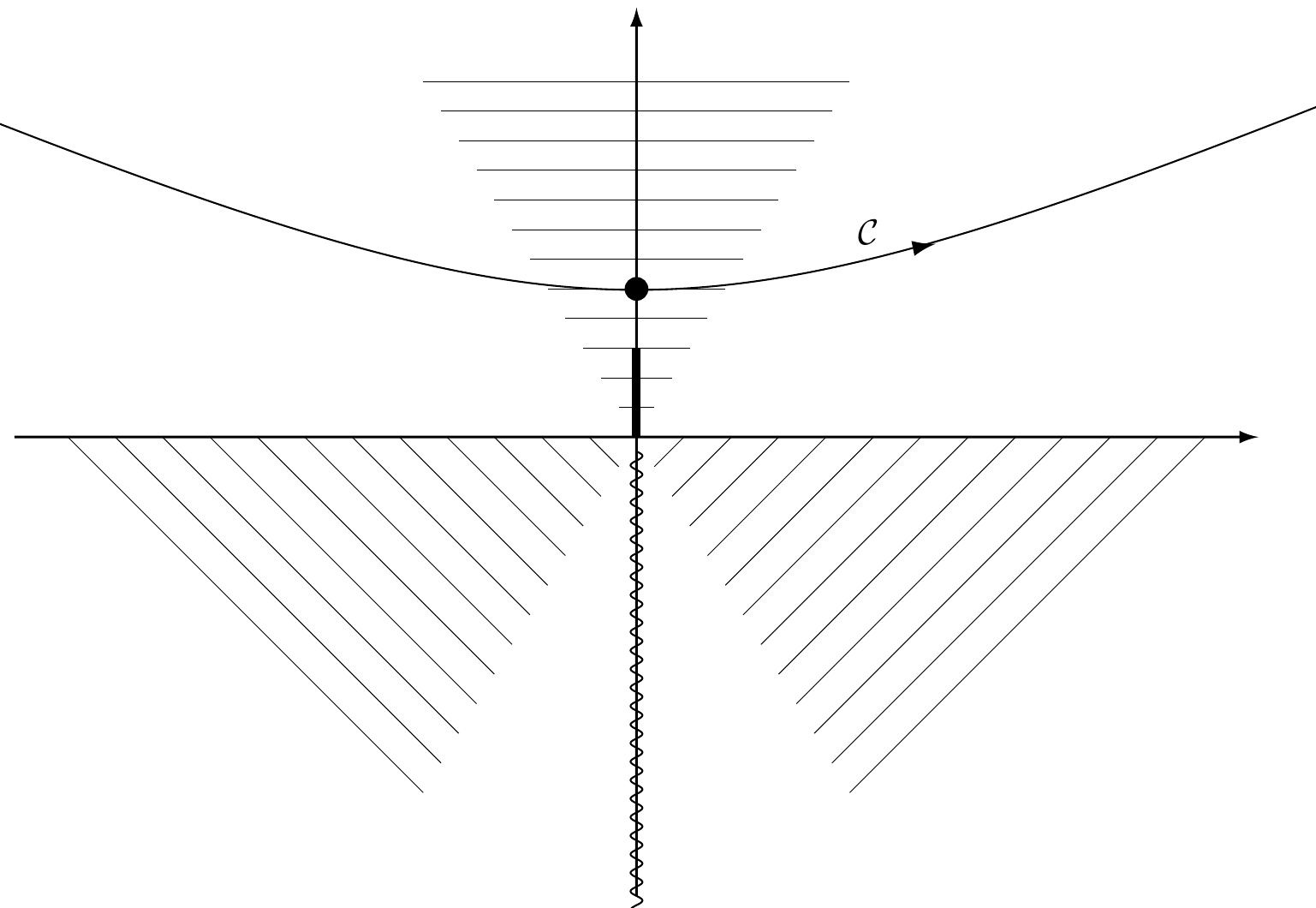}
\end{center}
\caption{Representation of the 
contours in the $p$-plane to get $(\phi_{-\om}^{\rm out})^*$,  
the {\it out} mode of negative norm.  
The hatched regions are the asymptotically forbidden ones. The wavy line is the branch cut of $\ln(p)$, and the bold line is what must be added because of the mass. 
The left panel is valid for $x>0$, and the right one for $x<0$. }
\label{masscontours} 
\end{figure}

In the following, we construct the generalization of the decaying mode of Sec.~\ref{decayingSec} 
because this is enough to extract the Bogoliubov coefficients. 
To this end, we must choose a branch cut to define both the  
$\ln(p)$ and 
$\sqrt{m^2+p^2}$ appearing in $G_m$. These functions introduce three branching points, $p=0$, and $p=\pm im$.
Here, we take the line $-i\mathbb R^+$ extended until $im$ to be the branch cut, 
as shown on Fig.~\ref{masscontours}. 
Hence in the limit $m \to 0$ we recover what we did in the body of the paper, see Fig.~\ref{contours}. 
To compute  \Eqref{massmode}, we proceed as in Sec.\ref{decayingSec}.

When $x>0$, the introduction of a mass 
does not alter the discussion of Sec.~\ref{decayingSec} since the saddle 
at $p_s = i\Lambda \sqrt{2\kappa x}$ is well 
above the singularity at $im$ for $x$ sufficiently far away from the horizon. Hence
\be
\varphi_{\omega,m}^{\mathcal C} = e^{iG_m(p_s)} \times  \varphi_{\omega}^{\downarrow} \times \left( e^{-i\frac{\pi}2} \right),
\ee
where $\varphi_{\omega}^{\downarrow}$ is the {\it massless dispersive} decaying mode of \Eqref{decmode}. 
Moreover, we have $G(p_s) \approx 0$
because $p_s \gg m$ in the region of interest. For  $x>0$, 
the mode is thus 
rapidly decaying, on a scale $\kappa x\sim (\kappa/\Lambda)^{\frac23}$,
and in the relativistic limit ({\it i.e.} $\Lambda \to \infty$) it vanishes. 
Therefore, as in the massless case, 
this mode is proportional to
$(\phi_{-\om, m}^{\rm out})^*$, the negative norm {\it out} going mode. 
Indeed, if it were containing a small amount of the positive norm {\it out} going mode 
$\phi_{\om, m}^{\rm out}$, 
it would oscillate on the right side of the horizon 
until $\kappa x\sim (\om/m)^2 \gg (\kappa/\Lambda)^{\frac23}$ which gives the location of the turning point where
the mode is reflected due to its mass, or due to a perpendicular momentum~\cite{TJRPBHententropy,Boom}. 

For $x<0$, using the analytic properties of $\tilde \varphi_\om (p)$, we  deform the contour $\mathcal C$ into the union of $\mathcal C_1$, $\mathcal C_2$ and $\mathcal C_3$ shown in Fig~\ref{masscontours}. 
On $\mathcal C_1$ and $\mathcal C_2$, there are two saddle points 
at $p = \pm \Lambda \sqrt{\kappa |x|}$ that describe the high momentum 
incoming modes, as in Sec.~\ref{decayingSec}. Their contribution is 
\be
\label{alpbet}
\varphi^{\mathcal C_1 \cup \mathcal C_2}_{\omega,m} = ( e^{\frac{\omega \pi}{\kappa}} e^{i\frac{3\pi}4} ) \times\left(\varphi_{-\omega}^{\rm in}\right)^* 
+  e^{i\frac{\pi}4}\times\varphi_\omega^{\rm in}, 
\ee
where $\varphi_{\pm \omega}^{\rm in}$ are the {massless} dispersive {\it in} modes of \Eqref{plusin} and \Eqref{minusin}. 
Moreover, the saddle point approximation is controlled by the parameter $\Delta(x)$ of \Eqref{Delta}, irrespectively of the mass $m \ll \Lambda$. 
Hence, as expected, the high momentum contributions of the {\it out} mode are mass independent. 
Along $\mathcal C_3$ instead, 
we perform a strong limit $\Lambda \to \infty$ as in \Eqref{gammalike},
and we get
\be
\varphi^{\mathcal C_3}_{\omega,m}(x) = \frac1{\sqrt{4\pi \kappa}} \int_{\mathcal C_3} 
p^{-i\frac{\omega}{\kappa}-1} e^{i (x p + G_m(p) )} \left(\frac{p}{\sqrt{p^2 + m^2}}\right)^{\frac12}
\frac{dp}{\sqrt{2\pi}} \label{C2},
\ee
which is a {\it massive relativistic} mode of negative norm. 
Up to a complex amplitude $A_\om$, it gives 
$(\varphi_{-\om, \, m}^{\rm out})^*$,
the low momentum out branch of the globally defined mode $(\phi_{-\om, \, m}^{\rm out})^*$.
In brief, the mode obtained with the contour $\mathcal C$ is 
$A_\om \left(\phi_{-\om,m}^{\rm out}\right)^*$: For all $\om$, it decays for $x > 0$, 
and on the left side, it contains three WKB branches 
\be
\label{concl}
A_\om \times \left(\phi_{-\om,m}^{\rm out}\right)^*
 = \left( \frac{\alpha_{\om, \, m}^*}{ \tilde 
\beta_{\om, \, m}^*} e^{i\frac{\pi}4} 
\right) \times\left(\varphi_{-\omega}^{\rm in}\right)^* 
+  e^{i\frac{\pi}4}\times
\varphi_\omega^{\rm in} + A_\om \times \left(\varphi_{-\om,m}^{\rm out}\right)^*.
\ee
Since $\left(\varphi_{-\omega}^{\rm in}\right)^* $ and $\varphi_\omega^{\rm in}$ are normalized
and have opposite norms, their relative coefficient furnishes the ratio of the {\it near horizon} Bogoliubov coefficients 
$\vert \tilde \beta_{\om, \, m}/\alpha_{\om, \, m}\vert$.~\footnote{\label{ftn}We have used the same 
conventions as in the massless case, see \Eqref{BogHR}, 
because the relevant part of the scattering matrix is still $2 \times 2$, for all values of $\om$.
This is non trivial because the dimensionality of the asymptotic modes is different from that of
the massless case. Indeed, when there is a mass, the left going mode $\phi_\om^{\rm left}$ only
exists above a critical frequency $\om_m^{\rm as}= m (1 - (v^{R}_{\rm as})^2)^{1/2}$ 
where $v^R_{\rm as}$ is 
the asymptotic velocity on the right side. However, as later discussed in this Appendix, the mode mixing with $\phi_\om^{\rm left}$ 
does not affect the asymptotic Bogoliubov coefficients. Hence $\phi_\om^{\rm left}$  remains a spectator mode.}
From \Eqref{alpbet} and \Eqref{concl}, we obtain  
$\vert \tilde \beta_{\om, \, m}/\alpha_{\om, \, m}\vert = e^{- \pi \om/\kappa}$. 
It is independent of $m$ and 
has the standard relativistic value. 

We decided to study $\left(\phi_{-\om,m}^{\rm out}\right)^*$
in order to be able to discuss
the {\it asymptotic} Bogoliubov coefficients. So far indeed 
our calculation
is restricted to the near horizon regions 1.b of Fig.~\ref{regions}. 
Had we studied the positive norm mode $\phi_\om^{\rm out}$, 
we would have faced the propagation in the regions 2 and 3 on the right of the horizon. 
Below the critical frequency $\om_m^{\rm as}$ discussed in footnote~\ref{ftn},
the mode $\phi_\om^{\rm out}$ is completely reflected.
Above that threshold,  $\phi_\om^{\rm out}$ is partially reflected in a non universal manner that depends
on the actual profile $v(x)$. However since this scattering is elastic, as it mixes left and right moving
positive norm modes (these being the only ones present in that region), 
it does not affect the pair creation probabilities of asymptotic quanta, {\it i.e.} it fixes 
`greybody factors'. In other words, 
the propagation in the right region has no influence on 
 pair creation probabilities, 
both for $\om$ below and above $\om_m^{\rm as}$.

On the contrary the mode mixing on the left regions 2 and 3 will  
affect them because this mixing involves modes with norms of both signs. 
For instance, in Minkowski
space-time no asymptotic particle is created even though \Eqref{concl} applies near the Rindler
horizon described by \Eqref{modequ} with
\be
v(x) = -(1-2\kappa x)^{\frac12}. 
\ee
This implies that the far away scattering {\it undoes} the one occurring near the horizon, 
as can be verified by direct calculation, using Bessel functions.  
For the smooth and asymptotically constant profiles we consider in this paper, 
WKB modes nevertheless provide a controlled approximation on the entire left region. 
Indeed, the errors due to the high momentum modes are $m$ independent and small 
whenever $\Delta_{\rm p}^L$ of  \Eqref{DpL} obeys $\Delta_{\rm p}^L \gg 1$.
The error due to the low momentum mode can be bounded using the treatment of App.\ref{appWKB}. 
For the reflected modes, for $0 \leqslant \om < \om_m^{\rm as}$, one gets
\be
\left|\frac{\beta_\om - \beta_\om^{\rm exact}}{\beta_\om^{\rm exact}}\right| = O\left(\frac{\kappa}{\om} \right).
\ee
Because the error grows like $1/\om$ 
for $\om \to 0$, there is no guarantee that the extra mode mixing will not significantly 
affect the values of the coefficients given in \Eqref{concl}. Hence further investigation is necessary to understand 
under which circumstances the standard divergence of $| \beta_\om |^2$ in $\kappa/\om$ is recovered in the massive case.
We are currently examining this interesting question. Let us conclude by noticing that since the asymptotic value of the 
wave vector $k^{\rm out}_\om$ is finite for $\om \to 0$, this opens the possibility of having a phenomenon similar to
what we found when studying white holes in Sec.~\ref{WHSec}.


\begin{thebibliography}{99}
\bibitem{Hawk75}
S.W. Hawking, ``Particle Creation by Black Holes'', Comm.\ Math.\ Phys. {\bf 43} 199 (1975).

\bibitem{BCH72}
J. M. Bardeen, B. Carter, and S. W. Hawking, ``The four laws of black hole mechanics,'' Comm.\ Math.\ Phys. {\bf 31} 161 (1973).

\bibitem{Bek73}
J.D. Bekenstein, ``Black Holes and Entropy,'' Phys.\ Rev.\ D {\bf 7}, 2333 2346 (1973).

\bibitem{Bek74}
J.D. Bekenstein, ``Generalized second law of thermodynamics in black-hole physics,'' Phys.\ Rev.\ D {\bf 9}, 3292 3300 (1974).

\bibitem{CallanMald}
C.~G.~Callan and J.~M.~Maldacena, ``D-brane Approach to Black Hole Quantum Mechanics,'' Nucl.\ Phys.\  B {\bf 472} (1996) 591 [arXiv:hep-th/9602043].

\bibitem{TJthermo}
T. Jacobson, ``Thermodynamics of Spacetime: The Einstein Equation of State,'' Phys.\ Rev.\ Lett. {\bf 75} (1995) 1260 [arXiv:gr-qc/9504004v2].

\bibitem{Unruh81}
W. G. Unruh, ``Experimental black hole evaporation ?,'' Phys.\ Rev.\ Lett.\  {\bf 46}, 1351 (1981).

\bibitem{tHooft85}
G. 't Hooft, ``On the quantum structure of a black hole,'' Nuc.\ Phys.\ {\bf B256} (1985) 727-745.

\bibitem{Primer}
R. Brout, S. Massar, R. Parentani, P. Spindel, ``A Primer for Black Hole Quantum Physics,'' Phys. Rept. {\bf 260} 329-454 (1995) [arXiv:0710.4345v1 [gr-qc]].

\bibitem{TJ91}
T. Jacobson, ``Black-hole evaporation and ultrashort distances,'' Phys.\ Rev.\ D {\bf 44} 1731-1739 (1991).

\bibitem{TJ93}
T. Jacobson, ``Black Hole Evaporation in the Presence of a Short Distance Cutoff,'' Phys. Rev. D {\bf 48} (1993) 728-741 [arXiv:hep-th/9303103v1].

\bibitem{beyond07}
R. Parentani, ``Beyond the semi-classical description of black hole evaporation,'' Int. J. Theor. Phys. {\bf 41} (2002) 2175-2200 [arXiv:0704.2563v1 [hep-th]].

\bibitem{Unruh95}
W. G. Unruh, ``Sonic analogue of black holes and the effects of high frequencies on black hole evaporation,'' Phys.\ Rev.\ D  {\bf 51}, 2827 (1995).

\bibitem{CJ96}
S. Corley and T. Jacobson, ``Hawking Spectrum and High Frequency Dispersion,'' Phys.\ Rev.\  D {\bf 54} (1996) 1568-1586 [arXiv:hep-th/9601073v1].

\bibitem{Macher1}
J. Macher, R. Parentani, ``Black/White hole radiation from dispersive theories,'' Phys.\ Rev.\ D {\bf 79}, 124008 (2009) [arXiv:0903.2224v3 [hep-th]].

\bibitem{BMPS}
R. Brout, S. Massar, R. Parentani, P. Spindel, ``Hawking radiation without transPlanckian frequencies,'' Phys.\ Rev.\  D {\bf 52} (1995) 4559 [arXiv:hep-th/9506121].

\bibitem{Corley}
S. Corley, ``Computing the spectrum of black hole radiation in the presence of high frequency dispersion: an analytical approach,'' Phys.\ Rev.\  D {\bf 57} (1998) 6280 [arXiv:hep-th/9710075].

\bibitem{Tanaka99}
Y. Himemoto, T. Tanaka, ``Generalization of the model of Hawking radiation with modified high frequency dispersion relation,'' Phys. Rev. D {\bf 61} (2000) 064004 [arXiv:gr-qc/9904076v3].

\bibitem{SU}
R. Schutzhold, W. G. Unruh, ``On the Universality of the Hawking Effect,'' Phys.\ Rev.\ D  {\bf 71}, 024028 (2005) [arXiv:gr-qc/0408009v2].

\bibitem{Rivista05}
R. Balbinot, A. Fabbri, S. Fagnocchi and R. Parentani, ``Hawking radiation from acoustic black holes, short distance and back-reaction effects,'' Riv.\ Nuovo Cim.\  {\bf 28}, 1 (2005) [arXiv:gr-qc/0601079].

\bibitem{Constructing07}
R. Parentani, ``Constructing QFT wherein Lorentz Invariance is broken by dissipative effects in the UV,'' PoS {\bf QG-PH} (2007) 031 [arXiv:0709.3943 [hep-th]].

\bibitem{Macherthesis}
J. Macher, ``Brisure de l'invariance de Lorentz à haute énergie : conséquences pour l'inflation et le rayonnement des trous noirs,'' PhD thesis, Univ. Paris-Sud (2009)

\bibitem{From2010}
R. Parentani, ``From vacuum fluctuations across an event horizon to long distance correlations,'' Phys.\ Rev.\  D {\bf 82} (2010) 025008  [arXiv:1003.3625 [gr-qc]].

\bibitem{Mayoral2011}
C. Mayoral, A. Recati, A. Fabbri, R. Parentani, R. Balbinot, I. Carusotto,``Acoustic white holes in flowing atomic Bose-Einstein condensates,'' New J. Phys. 13:025007 (2011) [arXiv:1009.6196v1 [cond-mat.quant-gas]].

\bibitem{SilkePRL2010}
S. Weinfurtner, E. W. Tedford, M. C. J. Penrice, W. G. Unruh, G. A. Lawrence,``Measurement of stimulated Hawking emission in an analogue system,'' Phys. Rev. Lett. 106:021302 (2011) [arXiv:1008.1911v2 [gr-qc]].

\bibitem{FP2}
S. Finazzi, R. Parentani, ``Spectral properties of acoustic black hole radiation: broadening the horizon,'' [arXiv:1012.1556v1 [gr-qc]].

\bibitem{FP3}
S. Finazzi, R. Parentani, ``On the robustness of acoustic black hole spectra,'' to appear in the proceedings of the Spanish Relativity Meeting ERE2010 [arXiv:1102.1452v1 [gr-qc]].


\bibitem{MacherBEC}
J. Macher, R. Parentani, ``Black-hole radiation in Bose-Einstein condensates,'' Phys.\ Rev.\ A {\bf 80}, 043601 (2009) [arXiv:0905.3634v4 [cond-mat.quant-gas]].

\bibitem{TJ96}
T. Jacobson, ``On the Origin of the Outgoing Black Hole Modes,'' Phys. Rev. D {\bf 53} (1996) 7082-7088 [arXiv:hep-th/9601064v2].


\bibitem{AnalogueLivingReview}
C.~Barcelo, S.~Liberati, M.~Visser, ``Analogue gravity,'' Living Rev.\ Rel.\  {\bf 8}, 12 (2005) [arXiv:gr-qc/0505065].

\bibitem{River}
T. Jacobson, ``Trans-Planckian redshifts and the substance of the space-time river'', [arXiv:hep-th/0001085v2].

\bibitem{SU08}
R. Schutzhold, W. G. Unruh, ``On the origin of the particles in black hole evaporation,'' Phys.\ Rev.\ D  {\bf 78}, 041504(R) (2008) [arXiv:0804.1686v1 [gr-qc]].


\bibitem{Waldbook}
R. Wald, ``General Relativity,'' University of Chicago press (1984).

\bibitem{BirrelDavies}
N. Birrell, P. Davies, ``Quantum Fields in Curved Space,'' Cambridge Monographs on Mathematical Physics (1982).

\bibitem{Page76}
N. Page, ``Particle Emission Rates from a Black Hole: Massless Particles from an Uncharged, Nonrotating Hole,'' Phys.\ Rev.\ D  {\bf 13}, 198206 (1976).

\bibitem{aQuattro}
A. Coutant, S. Finazzi, S. Liberati, R. Parentani, {\it work in progress}.

\bibitem{AbramoSteg}
M. Abramowitz, I. Stegun, ``Handbook of Mathematical Functions with Formulas, Graphs, and Mathematical Tables,'' New York, Dover Plublications (1964).

\bibitem{FP1}
S. Finazzi, R. Parentani, ``Black hole lasers in Bose-Einstein condensates,'' New.\ J.\ Phys. {\bf 12}, 095015 (2010) [arXiv:1005.4024v2 [cond-mat.quant-gas]].

\bibitem{Gottfried}
K. Gottfried ``Quantum Mechanics: Fundamentals" W.A.Benjamin, Inc., New York (1966)

\bibitem{Olver}
F. Olver, ``Asymptotics and special functions," Academic press, New York (1974).

\bibitem{Carusotto2008}
I. Carusotto, S. Fagnocchi, A. Recati, R. Balbinot, A. Fabbri,``Numerical observation of Hawking radiation from acoustic black holes in atomic Bose-Einstein condensates,'' New J. Phys. 10:103001 (2008) [arXiv:0803.0507v2 [cond-mat.other]].

\bibitem{Balbinot2008}
R. Balbinot, A. Fabbri, S. Fagnocchi, A. Recati, I. Carusotto,``Non-local density correlations as signal of Hawking radiation in BEC acoustic black holes,'' Phys. Rev. A 78:021603 (2008) [arXiv:0711.4520v2 [cond-mat.other]].

\bibitem{UnruhCor}
R. Schutzhold, W. G. Unruh, ``On Quantum Correlations across the Black Hole Horizon,'' Phys.\ Rev.\ D  {\bf 81}, 124033 (2010) [arXiv:1002.1844v1 [gr-qc]].

\bibitem{Zapata2011}
I. Zapata, M. Albert, R. Parentani and F. Sols, ``Resonant Hawking radiation in Bose-Einstein condensates,'' to appear in New J. Phys. [arXiv:1103.2994v2 [cond-mat.quant-gas]]. 

\bibitem{MacherCosmo}
J. Macher, R. Parentani, ``Signatures of trans-Planckian dispersion in inflationary spectra,'' Phys.\ Rev.\ D {\bf 78}, 043522 (2008) [arXiv:0804.1920v3 [hep-th]].

\bibitem{Rousseaux}
G. Rousseaux, P. Maissa, C. Mathis, P. Coullet, T. Philbin, U. Leonhardt,``Horizon effects with surface waves on moving water,'' New J. Phys. 12 (2010) 095018 [arXiv:1004.5546v2 [gr-qc]].


\bibitem{CampoPar}
D. Campo, R. Parentani, ``Space-time correlations in inflationary spectra, a wave packet analysis,'' Phys.\ Rev.\ D {\bf 70} 105020 (2004) [arXiv:gr-qc/0312055v2]. 

\bibitem{DamourRuffini76}
T. Damour, R. Ruffini, ``Black-hole evaporation in the Klein-Sauter-Heisenberg-Euler formalism,'' Phys.\ Rev.\ D {\bf 14} 332 (1976). 

\bibitem{Unruh76}
W. G. Unruh, ``Notes on Black-hole evaporation,'' Phys.\ Rev.\ D {\bf 14} 870 (1976). 

\bibitem{scottthesis}
S. Robertson, ``Hawking Radiation in Dispersive Media,'' PhD thesis (2011) in University of St Andrews [arXiv:1106.1805v1].

\bibitem{BHlaser}
S. Corley and T. Jacobson, ``Black hole lasers,'' Phys.\ Rev.\  D {\bf 59} (1999) 124011 [arXiv:hep-th/9806203].

\bibitem{Ulflaser}
U. Leonhardt and T. G. Philbin, ``Black Hole Lasers Revisited,'' [arXiv:0803.0669 [gr-qc]].

\bibitem{lasermode}
A. Coutant and R. Parentani, ``Black hole lasers, a mode analysis,'' Phys. Rev. D {\bf 81}, (2010) 084042 [arXiv:0912.2755v2 [hep-th]].

\bibitem{GRouss}
 J. Nardin, G. Rousseaux, and P. Coullet, ``Wave-Current Interaction as a Spatial Dynamical System: Analogies with Rainbow and Black Hole Physics,'' Phys.\ Rev.\ Lett. {\bf 102},  1245041 (2009).

\bibitem{TJRPBHententropy}
T.~Jacobson and R.~Parentani, ``Black hole entanglement entropy regularized in a freely falling frame,'' Phys.\ Rev.\  D {\bf 76} (2007) 024006 [arXiv:hep-th/0703233].

\bibitem{Boom}
G.~Jannes, P.~Maissa, T.~G.~Philbin and G.~Rousseaux, ``Hawking radiation and the boomerang behaviour of massive modes near a horizon,'' Phys.\ Rev.\  D {\bf 83} (2011) 104028 [arXiv:1102.0689 [gr-qc]].


\end{thebibliography}
\end{document}